\documentclass{aastex}
\usepackage{emulateapj5}
\usepackage{graphicx}\usepackage{epsf}
\usepackage{onecolfloat}

\newcommand{\PNM}{P(N|M)}

\newcommand{\Mmin}{M_{\mathrm{min}}}

\newcommand{\hkpc}{h^{-1}\mathrm{kpc}}
\newcommand{\hMsun}{h^{-1}M_{\odot}}

\newcommand{\N}{\langle N\rangle}
\newcommand{\Ns}{\langle N_s\rangle}
\newcommand{\Nc}{\langle N_c\rangle}
\newcommand{\NM}{\langle N\rangle_M}
\newcommand{\NsM}{\langle N_s\rangle_M}

\newcommand{\NN}{\langle N(N-1)\rangle}
\newcommand{\NsNs}{\langle N_s(N_s-1)\rangle}
\newcommand{\NNM}{\langle N(N-1)\rangle_M}

\newcommand{\NNN}{\langle N(N-1)(N-2)\rangle}

\newcommand{\NsNsNs}{\langle N_s(N_s-1)(N_s-2)\rangle}

\newcommand{\xibar}{\bar{\xi}}

\def\h1{h^{-1}}

\def\LCDM{$\Lambda$CDM}

\def\hMsun{h^{-1}{\ }{\rm M_{\odot}}}
\def\hMpc{h^{-1}{\ }{\rm Mpc}}

\def\hkpc{h^{-1}{\ }{\rm kpc}}

\begin{document}


\twocolumn[

\title{The Dark Side of the Halo Occupation Distribution}

\author{Andrey V. Kravtsov\altaffilmark{1}, Andreas A. Berlind\altaffilmark{1,2}, 
Risa H. Wechsler\altaffilmark{1,3}, Anatoly A. Klypin\altaffilmark{4}, Stefan Gottl\"ober\altaffilmark{5}, Brandon Allgood\altaffilmark{6}, Joel R. Primack\altaffilmark{6}}

\begin{abstract}  
  We analyze the halo occupation distribution (HOD) and two-point
  correlation function of galaxy-size dark matter halos using
  high-resolution dissipationless simulations of the concordance flat
  $\Lambda$CDM model. The halo samples include both the host halos
  and the {\em subhalos}, distinct gravitationally-bound halos within
  the virialized regions of larger host systems.  We find that the HOD, the
  probability distribution for a halo of mass $M$ to host a number of
  subhalos $N$, is similar to that found in semi-analytic and
  $N$-body$+$gasdynamics studies.  Its first moment, $\NM$, has a
  complicated shape consisting of a step, a shoulder, and a power-law
  high-mass tail.  The HOD can be described by Poisson statistics at
  high halo masses but becomes sub-Poisson for $\NM\lesssim 4$.  We
  show that the HOD can be understood as a combination of the
  probability for a halo of mass $M$ to host a central galaxy and the
  probability to host a given number $N_s$ of satellite galaxies.  The
  former can be approximated by a step-like function, while the latter
  can be well approximated by a Poisson distribution, fully specified
  by its first moment. The first moment of the satellite HOD can be
  well described by a simple power law $\Ns\propto M^{\beta}$ with
  $\beta\approx 1$ for a wide range of number densities, redshifts,
  and different power spectrum normalizations. This formulation provides a simple
  but accurate model for the halo occupation distribution found in
  simulations. At $z=0$, the two-point correlation function (CF) of
  galactic halos can be well fit by a power law down to $\sim
  100h^{-1}\ \rm kpc$ with an amplitude and slope similar to those of
  observed galaxies.  The dependence of correlation amplitude on the
  number density of objects is in general agreement with results from the SDSS
  survey.  At redshifts $z \gtrsim 1$, we find significant departures
  from the power-law shape of the CF at small scales, where the CF
  steepens due to a more pronounced one-halo component. The departures
  from the power law may thus be easier to detect in high-redshift
  galaxy surveys than at the present-day epoch. They can be used to
  put useful constraints on the environments and formation of
  galaxies.  If the deviations are as strong as indicated by our
  results, the assumption of the single power law often used in
  observational analyses of high-redshift clustering is dangerous and
  is likely to bias the estimates of the correlation length and slope
  of the correlation function.

\end{abstract}


\keywords{cosmology: theory--galaxies: formation-- galaxies: halos--large-scale structure of universe}
]

\altaffiltext{1}{Dept. of Astronomy and Astrophysics,
       Center for Cosmological Physics,
       The University of Chicago, Chicago, IL 60637;
       {\tt andrey@oddjob.uchicago.edu, aberlind@orbital.uchicago.edu}}
\altaffiltext{2}{Center for Cosmology and Particle Physics,
       New York University, New York, NY 10003 ;}
\altaffiltext{3}{Michigan Center for Theoretical Physics, Physics Department, University of Michigan, Ann Arbor, MI 48109;
       {\tt wechsler@umich.edu}}
\altaffiltext{4}{Astronomy Department, New Mexico State University,
MSC 4500, P.O.Box 30001, Las Cruces, NM 88003;
       {\tt aklypin@nmsu.edu}}
\altaffiltext{5}{ Astrophysikalisches Institut Potsdam,
An der Sternwarte 16,
14482 Potsdam, Germany;
       {\tt sgottloeber@aip.de}}
\altaffiltext{6}{Physics Department, University of California, Santa Cruz, CA 95064
       {\tt allgood@physics.ucsc.edu, joel@scipp.ucsc.edu}}

\section{Introduction}
\label{sec:intro}

Understanding the processes that drive galaxy clustering has always
been one of the main goals of observational cosmology. In particular,
the physical explanation for the approximately power-law shape of the
galaxy two-point correlation function \citep[e.g.,][and references
therein]{peebles80} is still an open problem. High-resolution
cosmological simulations over the past decade have shown that on small
scales ($\lesssim 1-2$~Mpc) the correlation function of matter strongly
deviates from the power-law shape. The direct implication of this
result is that the spatial distribution of galaxies on small scales is
biased with respect to the overall distribution of matter in a
non-trivial scale-dependent way
\citep[e.g.,][]{klypin_etal96,jenkins_etal98}.  In view of this, it is
very interesting to understand whether the power-law shape of the
correlation function is a fortuitous coincidence or a consequence of
some fundamental physical process.

The physics of galaxy formation, which almost certainly plays a role
in determining how galaxies of different types and luminosities are
clustered, is complicated and still rather poorly understood.  Galaxy
mergers, gas cooling, and star formation are just a few of the many
processes that can potentially affect the clustering statistics of a
galaxy sample. Nevertheless, despite the apparent complexity, there is
evidence that gravitational dynamics alone may explain the basic
features of galaxy clustering, at least in the simple case of galaxies
selected above a luminosity or mass threshold.  Building on several
pioneering studies
\citep[e.g.,][]{carlberg91,brainerd_villumsen92,brainerd_villumsen94a,brainerd_villumsen94b,colin_etal97},
\citet{kravtsov_klypin99} and \citet{colin_etal99} used
high-resolution $N$-body simulations that resolved both isolated halos
and dark matter substructure within virialized halos to show that the
correlation function of galactic halos has a power-law shape with an
amplitude and slope similar to those measured in the APM galaxy
catalog \citep{baugh96}.  More recently, \citet{neyrinck_etal03}
showed that dark matter subhalos identified in a different set of
dissipationless simulations have a correlation function and power
spectrum that matches that of the galaxies in the PSCz survey.  

These results suggest that the spatial distribution of galaxies can be
explained to a large extent simply by associating galaxies brighter
than a certain luminosity threshold with dark matter halos more
massive than a certain mass corresponding to that threshold. 
In practice, however, we can expect a considerable band-dependent scatter 
between galaxy luminosity and halo mass. The scatter in general needs to be 
accounted for in the model.  

Although the power spectrum and correlation functions provide a
relatively simple and useful measure of galaxy clustering, the
implications for the physics of galaxy formation are often difficult
to extract using these statistics alone.  The halo occupation
distribution (HOD) formalism, developed during the last several years, is
a powerful theoretical framework for predicting and interpreting
galaxy clustering. The formalism describes the bias of a class of
galaxies using the probability $\PNM$ that a halo of virial mass $M$
contains $N$ such galaxies and additional prescriptions that specify
the relative distribution of galaxies and dark matter within halos.
If, as theoretical models seem to predict
\citep{bond_etal91,lemson_kauffmann99,berlind_etal03}, the HOD at a
fixed halo mass is statistically independent of the halo's large-scale
environment, this description of galaxy bias is essentially complete.
Given the HOD, as well as the halo population predicted by a
particular cosmological model, one can calculate any galaxy clustering
statistic at both linear and highly non-linear scales.  In addition, the
HOD can be more easily related to the physics of galaxy formation than
most other statistics.

Several aspects of the HOD model have been studied using semi-analytic
galaxy formation models
\citep{kauffmann_etal97,governato_etal98,jing_etal98,kauffmann_etal99a,kauffmann_etal99b,benson_etal00a,benson_etal00b,sheth_diaferio01,
  somerville_etal01,wechsler_etal01,berlind_etal03} and cosmological
gasdynamics simulations
\citep{white_etal01,yoshikawa_etal01,pearce_etal01,berlind_etal03}.
\citet{berlind_etal03} present a detailed comparison of the HOD in a
semi-analytic model and gasdynamics simulations. They find that,
despite radically different treatments of the cooling, star formation,
and stellar feedback in the two approaches to galaxy formation
modeling, for galaxy samples of the same space density the predicted
HODs are in almost perfect agreement. This result lends indirect
support to the idea that the HOD, and hence galaxy clustering, is
driven primarily by gravitational dynamics rather than by processes
such as cooling and star formation. It is therefore interesting to
study the HOD that is predicted in purely dissipationless cosmological
simulations.  The probability distribution, $\PNM$, in this case is
measuring the probability for an isolated halo of mass $M$ to contain
$N$ subhalos within its virial radius.  As the observational
constraints on the HOD and its evolution improve, the predictions of
the halo HOD can be compared to the HOD of galaxies in order to
determine to what extent gravity alone is responsible for galaxy
clustering.

In this paper we use high resolution dissipationless simulations of
the concordance $\Lambda$CDM model to study the HOD of dark matter
halos and its evolution.  The paper is organized as follows: in
\S~\ref{sec:sim} and \S~\ref{sec:haloid} we describe the simulations
and the halo identification algorithm that we use. In
\S~\ref{sec:samples} we describe the halo samples used in our
analyses.  In \S~\ref{sec:halomodel} we review the main features of
the HOD formalism and the associated halo model of dark matter
clustering.  In \S~\ref{sec:hod} we present HOD predictions for dark
matter substructure and in \S~\ref{sec:cf} we show the corresponding
predictions for the two-point correlation function.  In
\S~\ref{sec:discussion} and \S~\ref{sec:conclusions} we discuss and 
summarize our results.

\begin{table}[tb]
\label{tab:sim}
\caption{Simulation parameters}
\begin{center}
\small
\begin{tabular}{cccccc}
\tableline\tableline\\
\multicolumn{1}{c}{Name}&
\multicolumn{1}{c}{$\sigma_8$}&
\multicolumn{1}{c}{$L_{\rm box}$}  &
\multicolumn{1}{c}{$N_{\rm p}$}  &
\multicolumn{1}{c}{$m_{\rm p}$}  &
\multicolumn{1}{c}{$h_{\rm peak}$} 
\\
\multicolumn{1}{c}{}&
\multicolumn{1}{c}{}&
\multicolumn{1}{c}{$h^{-1}\rm Mpc$} &
\multicolumn{1}{c}{} &
\multicolumn{1}{c}{$h^{-1}\rm\ M_{\odot}$} & 
\multicolumn{1}{c}{$h^{-1}\rm\ kpc$}  
\\
\\
\tableline
\\
$\Lambda$CDM$_{60}$ & 1.0 & 60  & $256^3$ & $1.07\times 10^9$ & $1.9$\\
$\Lambda$CDM$_{80}$ & 0.75 & 80  & $512^3$ & $3.16\times 10^8$ & $1.2$\\
\\
\tableline
\end{tabular}
\end{center}
\label{tab:simparam}
\end{table}

\section{Simulations}
\label{sec:sim}

We analyze the halo occupation distribution and clustering in the
concordance flat {\LCDM} model: $\Omega_0=1-\Omega_{\Lambda}=0.3$,
$h=0.7$, where $\Omega_0$ and $\Omega_{\Lambda}$ are the present-day
matter and vacuum densities, and $h$ is the dimensionless Hubble
constant defined as $H_0\equiv 100h{\ }{\rm km\ s^{-1}\,Mpc^{-1}}$.
This model is consistent with recent observational constraints
\citep[e.g.,][]{spergel_etal03}. To study the effects of the power
spectrum normalization and resolution we consider two simulations of
the {\LCDM} cosmology. The first simulation followed the evolution of
$256^3\approx 1.67\times 10^7$ particles in a $60\hMpc\approx
85.71$~Mpc box and was normalized to $\sigma_8=1.0$, where $\sigma_8$
is the rms fluctuation in spheres of $8h^{-1}{\ }{\rm Mpc}$ comoving
radius.  This simulation was used previously to study the halo
clustering and bias by \citet{kravtsov_klypin99} and
\citet{colin_etal99} and we refer the reader to these papers for
further numerical details.  This simulation was also used to study
halo concentrations \citep{bullock_etal01b}, the specific angular
momentum distribution \citep{bullock_etal01}, and the accretion
history of halos \citep{wechsler_etal02}.
The second simulation followed the evolution of $512^3\approx
1.34\times 10^8$ particles in the same cosmology, but in a
$80\hMpc\approx 114.29$~Mpc box and with a power spectrum
normalization of $\sigma_8=0.75$. This normalization is suggested by
several recent measurements
\citep[e.g.,][]{borgani_etal01,pierpaoli_etal01,lahav_etal02,schuecker_etal03,jarvis_etal03}.

The simulations were run using the Adaptive Refinement Tree $N$-body
code \citep[ART;][]{kravtsov_etal97,kravtsov99}. The ART code reaches
high force resolution by refining all high-density regions with an
automated refinement algorithm.  The refinements are recursive: the
refined regions can also be refined, each subsequent refinement having
half of the previous level's cell size.  This creates an hierarchy of
refinement meshes of different resolution covering regions of
interest.  The criterion for refinement is the mass of particles per
cell. In the $\Lambda$CDM$_{60}$ the code refined an individual cell
only if the mass exceeded $n_{th}=5$ particles independent of the
refinement level. In terms of overdensity, this means that {\em all}
regions with overdensity higher than $\delta = n_{th}{\ 
}2^{3L}/\bar{n}$, where $\bar{n}$ is the average number density of
particles in the cube, were refined to the refinement level $L$. Thus, for
the $\Lambda$CDM$_{60}$ simulation , $\bar{n}$ is $1/8$.  The peak formal
dynamic range reached by the code in this simulation is $32,768$,
which corresponds to the peak formal resolution (the smallest grid 
cell) of $h_{\rm peak}=1.83\hkpc$; the actual
force resolution is $\approx 2h_{\rm peak}=3.7\hkpc$
\citep[see][]{kravtsov_etal97}.  In the higher-resolution
$\Lambda$CDM$_{80}$ simulations the refinement criterion was level-
and time-dependent. At the early stages of evolution ($a<0.65$) the
thresholds were set to 2, 3, and 4 particle masses for the zeroth,
first, and second and higher levels, respectively. At low redshifts,
$a>0.65$, the thresholds for these refinement levels were set to 6, 5, and 5
particle masses.  The lower thresholds at high redshifts were set to
ensure that collapse of small-mass halos is followed with higher
resolution. The maximum achieved level of refinement was $L_{\rm
  max}=8$, which corresponds to the comoving cell size of $1.22\hkpc$.
As a function of redshift the maximum level of refinement was equal to
$L_{\rm max}=6$ for $5<z<7$, $L_{\rm max}= 7$ for $1<z<5$, $L_{\rm
  max}\geq 8$ for $z<1$. The peak formal resolution was $h_{\rm peak}\leq
1.2\hkpc$ (physical).  The parameters of the simulations are
summarized in Table~\ref{tab:simparam}.

\begin{figure*}[tp]
\centerline{\epsfxsize3.5truein\epsffile{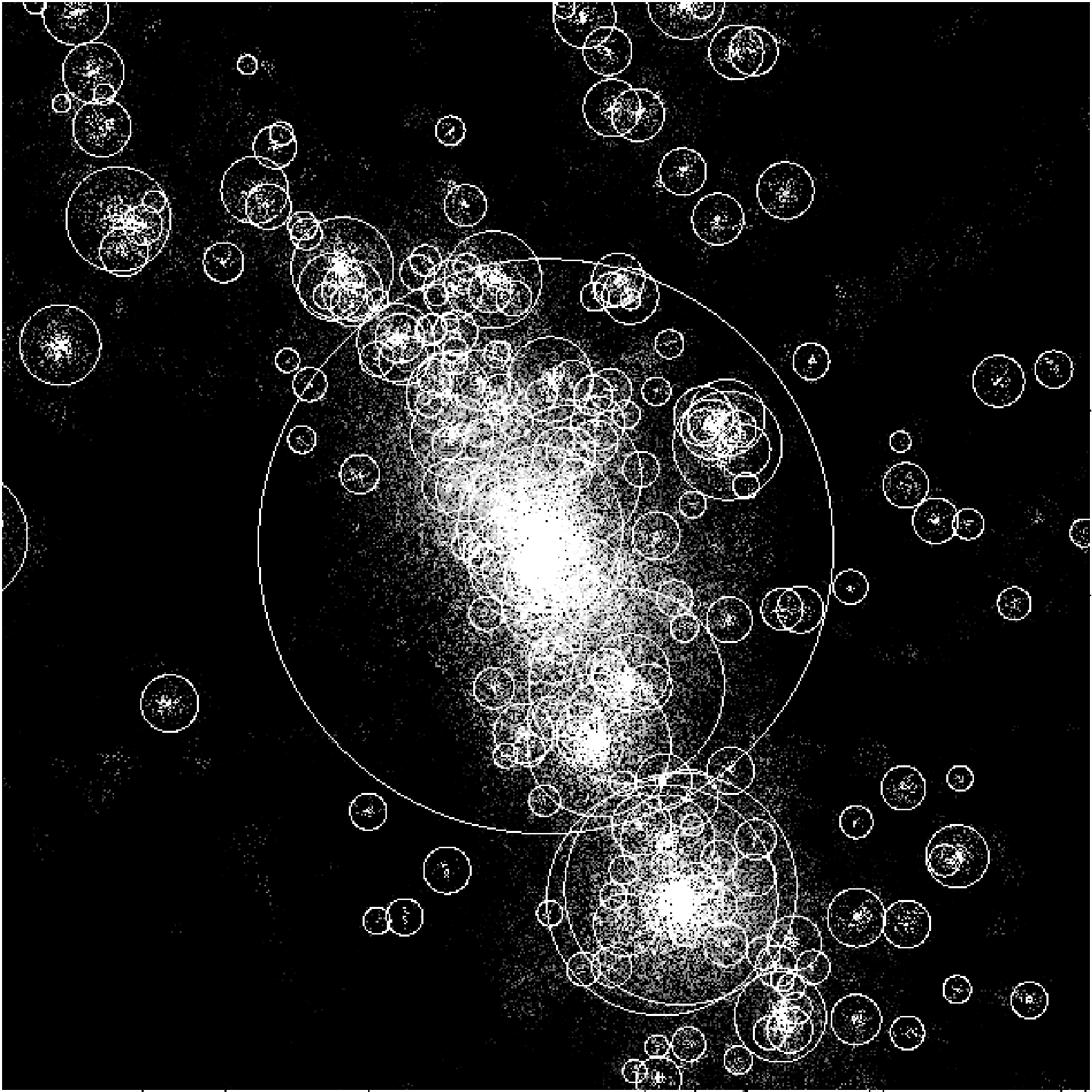}\hspace{0.5cm}\epsfxsize3.5truein\epsffile{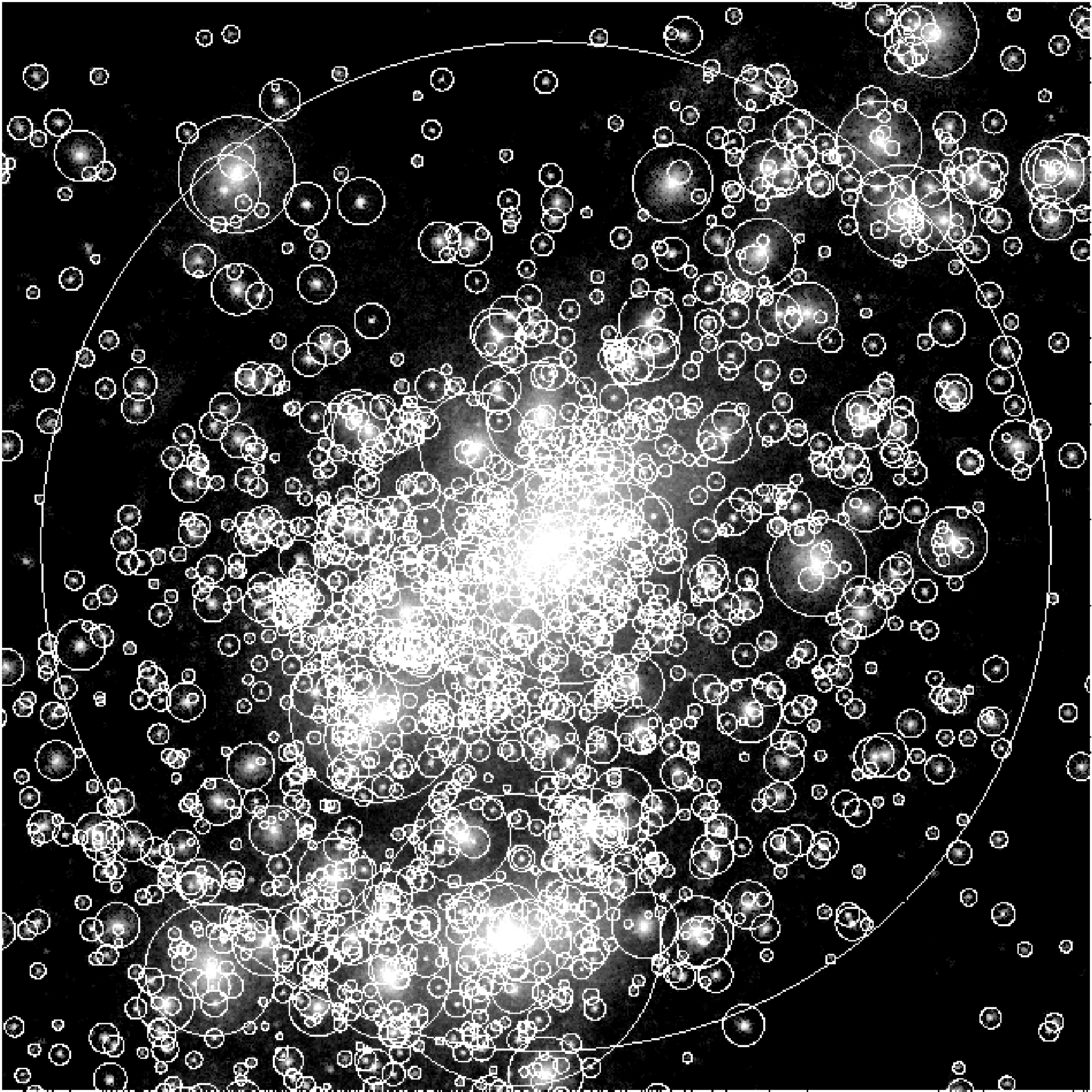}}
\caption{Distribution of dark matter particles (points) and dark matter halos
  (circles) identified by our halo finding algorithm centered on the
  most massive halo in the $\Lambda$CDM$_{80}$ simulation at $z=3$
  (left) and $z=0$ (right). The radius of the largest circle indicates
  the actual virial radius, $R_{180}$, of the most massive halo
  ($R_{180}=0.67h^{-1}$~comoving Mpc at $z=3$ and
  $R_{180}=2.1h^{-1}$~Mpc at $z=0$); the radii of all other halos are
  scaled using the halo' maximum circular velocities ($r_{\rm
    h}=0.65V_{\rm max}$~kpc with $V_{\rm max}$ in $\rm km\ s^{-1}$). 
\label{fig:dmh}}
\end{figure*}

\section{Halo identification}
\label{sec:haloid}

Identification of DM halos in the very high-density environments
of groups and clusters is a challenging problem. The goal
of this study is to investigate the halo occupation distribution and
clustering of the overall halo population. Therefore, we need to
identify both {\it host} halos with centers that do not lie 
within any larger virialized system and {\it subhalos} located within the
virial radii of larger systems. Below we use the terms {\em satellites}, 
{\em subhalos}, and {\em substructure} interchangeably.

To identify halos and the subhalos within them we use a variant of the
Bound Density Maxima (BDM) halo finding algorithm \citet[][hereafter
KGKK]{klypin_etal99}. We start by calculating the local overdensity at
each particle position using the SPH smoothing kernel\footnote{To
  calculate the density we use the publicly available code {\tt
    smooth}: {\tt
    http://www-hpcc.astro.washington.edu/tools/tools.html}} of 24
particles. The number of kernel particles roughly corresponds to the
lowest halo mass that we hope to identify. We then sort particles
according to their overdensity and use all particles with $\delta \geq
\delta_{\rm min}=2000$ as potential halo centers.  The specific value
of $\delta_{\rm min}$ was chosen after experimentation to ensure
completeness of the halo catalogs on the one hand while maximizing
the efficiency of the halo finder. 

Starting with the highest overdensity particle, we surround each
potential center by a sphere of radius $r_{\rm find}=50h^{-1}\ \rm
kpc$ and exclude all particles within this sphere from further center
search.  The search radius is defined by the size of smallest systems
we aim to identify. We checked that results do not change if this
radius is decreased by a factor of two. After all potential centers
are identified, we analyze the density distribution and velocities of
surrounding particles to test whether the center corresponds to a
gravitationally bound clump. Specifically, we construct density,
circular velocity, and velocity dispersion profiles around each center
and iteratively remove unbound particles using the procedure outlined
in \citet{klypin_etal99b}. We then construct final profiles using only
bound particles and use them to calculate properties of halos such as
maximum circular velocity $V_{\rm max}$, mass $M$, etc.

The virial radius is meaningless for the subhalos within a larger host
as their outer layers are tidally stripped. The definition of the
outer boundary of a subhalo and its mass are thus somewhat ambiguous.
We adopt the {\rm truncation radius}, $r_{\rm t}$, at which the
logarithmic slope of the density profile constructed from the bound
particles becomes larger than $-0.5$, as we do not expect the density
profile of the CDM halos to be flatter than this slope. Empirically,
this definition roughly corresponds to the radius at which the density of
the gravitationally bound particles is equal to the background host
halo density, albeit with a large scatter. For some halos $r_{\rm t}$
is larger than their virial radius. In this case, we set $r_{\rm
  t}=R_{\rm vir}$. For each halo we also construct the circular
velocity profile $V_{\rm circ}(r)=\sqrt{GM(<r)/r}$ and compute the
maximum circular velocity $V_{\rm max}$. 

Figure~\ref{fig:dmh} shows the particle distribution in the most
massive halos identified at $z=0$ and $z=3$ along with the halos
(circles) identified by the halo finder. The particles are color-coded
on a grey scale according to the logarithm of their density to enhance
visibility of substructure clumps. The radius of circles is
proportional to the halo's $V_{\rm max}$.  The figure shows that the
algorithm is efficient in identifying substructure down to small
masses. Note that the smallest halos plotted in Fig.~\ref{fig:dmh}
have masses smaller than our completeness limit of $\approx 50$
particles. This approximate limit corresponds to the mass below which
cumulative mass and velocity functions start to flatten
significantly. In the following analysis, we will consider only halos 
with masses $M>50m_{\rm p}$ (corresponding to $1.6 \times 10^9$ $\hMsun$ and 
$5.4\times 10^{10}$ $\hMsun$ in the $\Lambda$CDM$_{80}$ 
and $\Lambda$CDM$_{60}$ boxes respectively).

To classify the halos, we calculate the formal boundary of each object
as the radius corresponding to the enclosed overdensity of 180 with
respect to the mean density around its center. The halos whose center
is located within the boundary of a larger mass halo we call {\it
  subhalos} or {\it satellites}. The halos that are not classified as
satellites are identified as {\it host} halos. Note that the center of
a host halo is not considered to be a subhalo. Thus, host halos may or 
may not contain any subhalos with circular velocity above the threshold 
of a given sample. The host centers, however, are included
in clustering statistics because we assume that each host harbors 
a {\it central} galaxy at its center. Therefore, the total sample of galactic
halos contains central and satellite galaxies. The former have positions
and maximum circular velocities of their host halos, while the latter
have positions and $V_{\rm max}$ of subhalos. 

In the observed universe, the analogy is simple. The Milky Way, for
example, would be the central galaxy in a host halo of mass $M_{\rm
  h}\sim 10^{12}h^{-1}\ \rm M_{\odot}$ because its center is not
within any larger virialized system.\footnote{Note that the Local
  Group is not virialized and the Milky Way and Andromeda reside in two
  independent host halos.} The host halo of the Galaxy contains a
number of satellites, which would or would not be included in a galaxy
sample depending on how deep the sample is. In a rich cluster, the brightest 
cluster galaxy that typically resides near the cluster center would be 
associated with the cluster host halo in our terminology. All other 
galaxies within the virial radius of the cluster would be considered
``satellites'' associated with subhalos. 

\section{Halo Samples}
\label{sec:samples}

To construct a halo catalog, we have to define selection criteria based
on particular halo properties.  Halo mass is usually used to define
halo catalogs (e.g., a catalog can be constructed by selecting all
halos in a given mass range).  However, the mass and radius are very
poorly defined for the satellite halos due to tidal stripping which
alters a halo's mass and physical extent (see KGKK). Therefore, we will
use maximum circular velocity $V_{\rm max}$ as a proxy for the halo mass.  This
allows us to avoid complications related to the mass and radius
determination for satellite halos. Moreover, when a halo gets stripped 
$V_{\rm max}$ changes less dramatically than the mass, and is therefore a
more robust ``label'' of the halo. For isolated halos, $V_{\rm max}$ and the
halo's virial mass are directly related. For the suhalos $V_{\rm max}$
will experience secular decrease but at a relatively slow rate.

Instead of selecting objects in a given range of $V_{\rm max}$, at
each epoch we will select objects of a fixed set of number densities
corresponding to (redshift-dependent) thresholds in maximum circular
velocity: $n_i(>V_{\rm max})$.  Note that the number density here
includes all the centers of the isolated host halos and the subhalos
within the hosts (see eq.~\ref{eq:nhalo} in \S~\ref{sec:hodmodel}).

The threshold selection is somewhat
arbitrary, except that the limited box size puts a lower limit on the
number densities we can consider. The completeness limit of our
catalogs imposes an upper limit on the number densities we can
consider.  The number densities probed in this simulation span a
representative range of galaxy number densities.  We chose to focus on
a set of number densities corresponding to a representative set of
luminosity cuts for SDSS galaxies. Namely, we use the Schechter fit 
to the SDSS $r$-band
luminosity function presented by \citet{blanton_etal03} and select the
set of galaxy number densities corresponding to the absolute magnitude
thresholds $M_r=-16$, $-18$, $-19$, $-20$, and $-21$ (the magnitudes 
quoted are $M-5\log h$). The number
densities and corresponding numbers of galactic halos in the analyzed
simulations are listed in Table~\ref{tab:thresholds}.
Note that the median redshift of galaxies in the sample of 
\citet{blanton_etal03} is $z=0.1$, while we use simulation outputs
at $z=0$. However, using halos at the
$z=0.1$ output instead of z=0.0 results in only 2\% change in the values
of threshold $V_{\rm max}$.

\begin{figure}[t]
\vspace{-0.5cm}
\centerline{\epsfxsize=3.9truein\epsffile{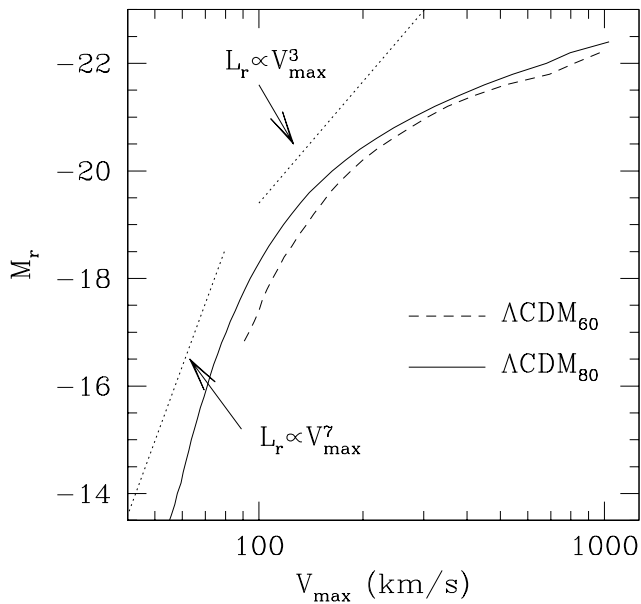}}
\vspace{-0.5cm}
\caption{The maximum circular velocity of halos in the $60\hMpc$ ({\it dashed} line) 
  and $80\hMpc$ ({\it solid} line) $\Lambda$CDM simulations vs. the
  $r$-band absolute magnitude of the SDSS galaxies of the same number
  density. The curves are obtained by matching the cumulative 
velocity functions $n(>V_{\rm max})$ (at $z=0$) to the SDSS luminosity 
function $n(<M_r)$. The dashed lines show the power-law luminosity--circular
velocity relation, $L_r\propto V_{\rm max}^{a}$, for $a=7$ and $a=3$.
\label{fig:mrvm}}
\vspace{-0.5cm}
\end{figure}

Figure~\ref{fig:mrvm} shows the maximum circular velocity of the halos
in our simulations with the same number density as the SDSS galaxies
with a given $M_r$.  For comparison the dotted lines show the power
law luminosity--circular velocity relation, $L_r\propto V_{\rm
  max}^{a}$, with $a=7$ and $a=3$.  Note that the relation does not
have a power law form at any circular velocity.  For $100<V_{\rm
  max}<200\rm\ km\ s^{-1}$, the slope is $a\approx 3$, while for
$V_{\rm max}<100\rm\ km\ s^{-1}$ the slope is much steeper: $a\approx
7$.  Such steepening is of course required to match the
shallow faint-end of the galaxy luminosity function with the
relatively steep circular velocity function. At $V_{\rm max}\gtrsim
300\rm  km\,s^{-1}$, the relation becomes shallow because the number
density of halos at these masses is dominated by the central
``galaxies'' which are assigned maximum circular velocity of their
group- or cluster-size host halo.  The overall shape of the
$M_r-V_{\rm max}$ relation thus likely reflects the non-monotonic
dependence of the mass-to-light ratio $M_{\rm h}/L$ on the host mass
\citep[e.g.,][]{benson_etal00a,vandenbosch_etal03}. The detailed comparisons with the
observed Tully-Fisher, however, require more detailed modeling which
takes into account effects of baryon cooling on $V_{\rm max}$. At
$V_{\rm max}>300\rm\ km\ s^{-1}$, the maximum circular velocity is
measured for the host group and cluster-size systems, rather than for
the central object, as it is impossible to unambiguously separate the
central object from the host group in dissipationless simulations.
\begin{figure*}[tp]
\centerline{\epsfxsize4truein\epsffile{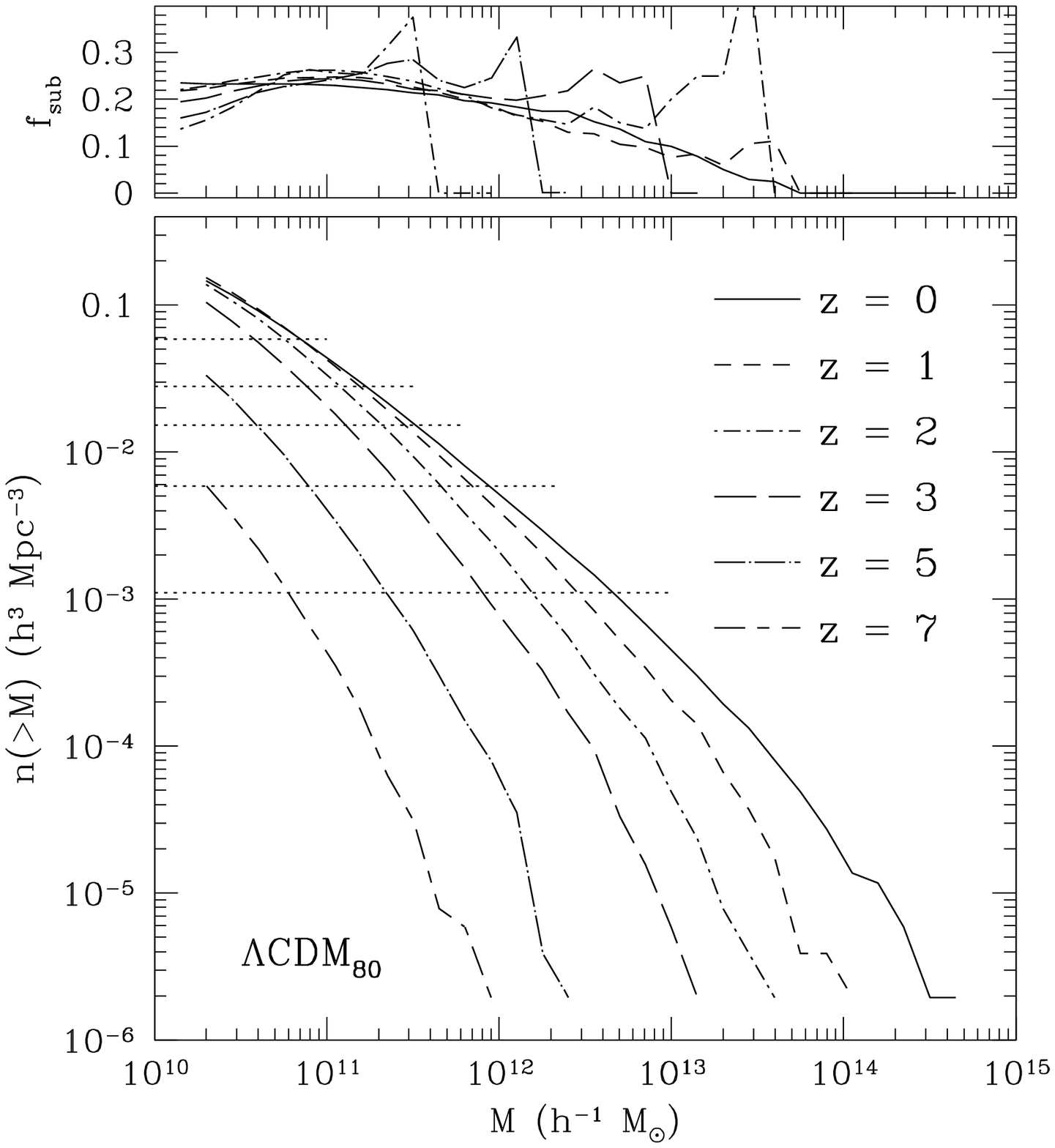}\vspace{-0.25cm}\hspace{-1cm}\epsfxsize4truein\epsffile{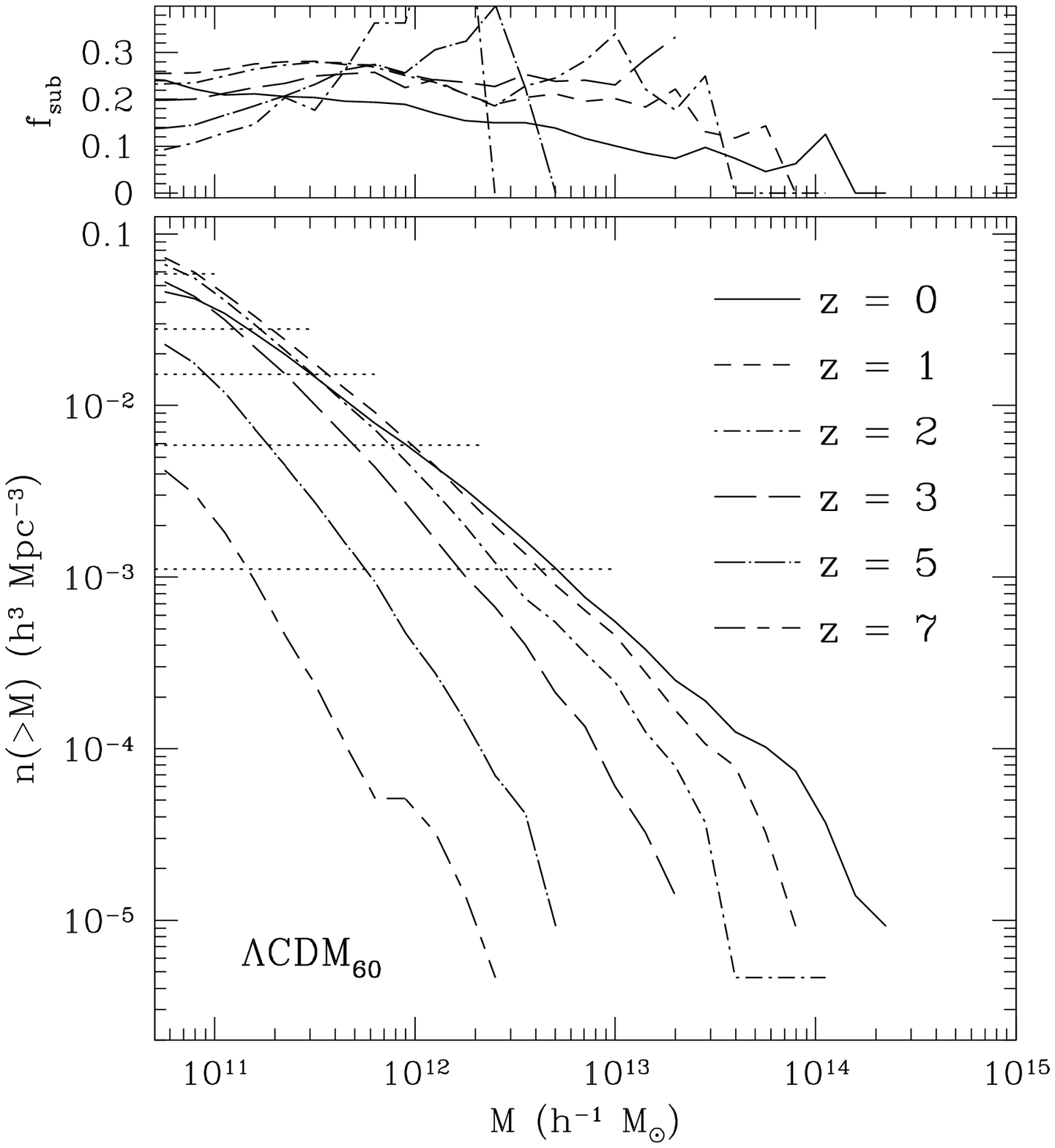}}
\caption{{\it Bottom panels:} cumulative mass functions of the halo samples (left panel: $\Lambda$CDM$_{80}$, right panel: $\Lambda$CDM$_{60}$) used
  in our analysis at different redshifts. Note that the number density
  here includes all the centers of the isolated host halos and the
  subhalos located within the hosts (see eq.~\ref{eq:nhalo} in
  \S~\ref{sec:hodmodel}).  The horizontal dotted lines indicate the
  number density thresholds adopted in our analysis. The curves were
  plotted down to the minimum halo mass of 50 particles.
{\it Top panels:} the fraction of halos with masses larger than $M$ 
classifed as subhalos: $f_{\rm sub}=(n - n_{\rm host})/n$.
\label{fig:vmf}}
\end{figure*}

Figure~\ref{fig:vmf} shows the cumulative mass functions for the
$\Lambda$CDM$_{60}$ and $\Lambda$CDM$_{80}$ simulations at the
analyzed epochs. The functions include both centers of the host halos
and subhalos and are only plotted down to the mass corresponding to
$50$ particles, below which our halo catalogs are incomplete.  The
horizontal dotted lines show the number density thresholds adopted for
the subsequent analysis.  Note that at high redshifts the halo
catalogs are incomplete at the highest number densities.  The figure
also shows the fraction of $n(>M)$ that is in objects classified as
subhalos. Note that in our samples of galaxy-size halos, about
$15-25\%$ of all the halos are in subhalos at any epoch.  This is
comparable to the observed fraction of $\sim 20\%$ of galaxies located
in groups and clusters.

\begin{table}[tb]
\caption{Number density thresholds}
\begin{center}
\small
\begin{tabular}{ccrr}
\tableline\tableline\\
\multicolumn{1}{c}{$n$}&
\multicolumn{1}{c}{SDSS $M_r$}&
\multicolumn{1}{c}{$N_{\rm halo}^{80}$}&
\multicolumn{1}{c}{$N_{\rm halo}^{60}$}  
\\
\multicolumn{1}{c}{$h^3\ \rm Mpc^{-3}$}&
\multicolumn{1}{c}{}&
\multicolumn{1}{c}{}&
\multicolumn{1}{c}{} 
\\
\\
\tableline
\\
$5.86\times 10^{-2}$ & -16 & 30003 & 12657\\ 
$2.79\times 10^{-2}$ & -18 & 14285 & 6026 \\
$1.52\times 10^{-2}$ & -19 &  7782 & 3282 \\
$5.89\times 10^{-3}$ & -20 &  3016 & 1272 \\
$1.11\times 10^{-3}$ & -21 &  568  & 240  \\
\\
\tableline
\end{tabular}
\end{center}
\label{tab:thresholds}
\end{table}

\section{The Halo Model}
\label{sec:halomodel}

The description of different elements of the halo model can be found
in several recent papers
\citep[e.g.,][]{seljak00,ma_fry00,peacock_smith00,scoccimarro_etal01,sheth_etal01b,sheth_etal01a,berlind_weinberg02,cooray_sheth02,yang_etal03}.
The model has quickly proven to be a very convenient analytic formalism
for predicting and interpreting the nonlinear clustering of dark
matter and galaxies.  The main idea behind the model is that all dark
matter is bound up in halos that have well-understood properties.  The
dark matter distribution is then fully specified by 1) the halo mass
function, 2) the linear bias of halos as a function of halo mass, and 3) the
radial density profiles of halos as a function of halo mass.  These
three elements have been relatively well studied using N-body
simulations and they can be computed analytically, given a
cosmological model.  In order to specify the galaxy distribution, two
additional ingredients are required: 4) the probability distribution
$\PNM$ that a halo of virial mass $M$ contains $N$ galaxies and 5) the
relative distribution of galaxies and dark matter within halos.  These
last two elements are called the halo occupation distribution (HOD)
and they are only now becoming the focus of much attention both
theoretically and observationally (\citealt{berlind_etal03} and
references therein).  $\PNM$ is the most important piece of the HOD in
terms of its effect on galaxy clustering and it is the main focus of
this study.  We refer to this element when we use the term HOD henceforth.

In the halo model the two-point correlation function of the galaxy distribution
is a sum of two terms: the ``1-halo'' term due to galaxy pairs within a
single halo and the ``2-halo'' term due to pairs in separate distinct
halos:
\begin{equation}
\xi_{\rm gg}(r)=\xi^{\rm 1h}_{\rm gg}(r)+\xi^{\rm 2h}_{\rm gg}(r) + 1.
\label{eq:xigg}
\end{equation}
At scales larger than the virial diameter of the largest halos, all
pairs consist of galaxies in separate halos ($\xi^{\rm 1h}\ll \xi^{\rm 2h}$), 
while at smaller scales most pairs consist of galaxies within the same halo
($\xi^{\rm 1h}\gg\xi^{\rm 2h}$).  The two terms are given by
\begin{equation}
1+\xi^{\rm 1h}_{\rm gg}(r)=\frac{1}{2}\,\bar{n}_{\rm g}^{-2}\int n(M)\NNM \lambda(r|M)\, dM ;
\label{eq:xi1h}
\end{equation}
\begin{eqnarray}
\xi^{\rm 2h}_{\rm gg}(r)&=& \xi^{\rm lin}_{\rm mm}(r)\,\,\bar{n}_{\rm g}^{-2}\int
n(M_1)b_h(M_1)\N_{M_1}\, dM_1 \nonumber\\
&  & \int n(M_2)b_h(M_2)\N_{M_2} \lambda(r|M_1,M_2)\, dM_2
\label{eq:xi2h}
\end{eqnarray}
where $\bar{n}_{\rm g}$ is the mean number density of galaxies in
the sample, $n(M)$ is the halo mass function, $b_h(M)$ is the
large-scale linear bias of halos, $\lambda(r|M)$ is the convolution of
the radial profile of galaxies within halos with itself,
$\lambda(r|M_1,M_2)$ is the convolution of two different radial
profiles, and $\xi^{\rm lin}_{\rm mm}(r)$ is the linear dark matter
correlation function \citep[also
see][]{sheth_etal01a,sheth_etal01b,berlind_weinberg02}. The
integration is over the mass limit corresponding to the galaxy sample
under consideration. 

On large scales, the 2-halo term reduces to $\xi^{\rm 2h}_{\rm
  gg}(r)=b^2\xi^{\rm lin}_{\rm mm}(r)$, where $b$ is the large-scale
bias factor of galaxies.  Equations~\ref{eq:xi1h} and~\ref{eq:xi2h}
represent the halo model in its most basic form, but variations do
exist \citep[see, e.g.,][]{zehavi_etal03,magliocchetti_porciani03}.
The halo model is most easily applied in Fourier space because the
calculation of the correlation function in real space involves
convolutions, which turn into multiplications in Fourier space.  For
our purposes, however, it suffices to express galaxy clustering in
terms of the real-space correlation function.

Equations ~\ref{eq:xi1h} and \ref{eq:xi2h} show that $\xi^{\rm 2h}$ 
depends on the average number of galaxies per halo of a given mass, 
\begin{equation}
\NM =\sum_{N} N P(N\vert M),
\label{eq:Nave}
\end{equation}
while $\xi^{\rm 1h}$ depends on the second moment of $\PNM$
\begin{equation}
\NNM =\sum_{N} N(N-1) P(N\vert M).
\label{eq:NNave}
\end{equation}
Higher order correlations depend on the higher order moments of
$\PNM$.  Both the mean $\NM$ and the shape of $P(N\vert M)$ are thus
key components of the halo model.

For galaxy samples defined by a minimum luminosity threshold, the mean
halo occupation $\NM$ is usually assumed to be a power law at high
halo masses ($M\gtrsim 10^{13}$ $\hMsun$).  This is supported both by
theoretical models (\citealt{berlind_etal03} and references therein)
and by halo model fits to observational data
\citep[e.g.,][]{scoccimarro_etal01,zehavi_etal03,magliocchetti_porciani03}.
At low halo masses, $\NM$ is expected to reach a plateau $\N\sim 1$,
where each halo contains on average only one galaxy, and then cut off
below a minimum mass threshold.  

An alternative approach is to assume the existence of two separate
galaxy populations: central halo galaxies (zero or one per halo) and
satellite galaxies within halos. These populations can then be
modeled separately \citep[e.g.,][]{guzik_seljak02}.  In particular,
in the most simple case the HOD of central galaxies can be modeled as
a step function $\Nc_M=1$ with $\Nc_M=0$ for $M<M_{\rm min}$, while the
HOD of the satellite galaxies can be modeled as a power law,
$\NsM\propto M^{\beta}$.  The motivation for the separation of central
and satellites galaxies comes partly from the analysis of hydrodynamic
simulations \citep{berlind_etal03}, and partly from studies of central 
bright elliptical galaxies in groups and clusters, which are often 
considered as a separate population from the rest of galaxies. As we 
show below, such a separation greatly simplifies the HOD analysis.

Several simple distributions have been considered for the shape of the
HOD.  The {\em Poisson} distribution is fully specified by its first
moment $\N$, as the high-order moments are simply 
\begin{equation}
\langle N(N-1)...(N-j)\rangle=\N^{j+1}.
\label{eq:hopoisson}
\end{equation}
For the {\em nearest integer} distribution with $N_l\lesssim\N<N_l+1$ (where $N_l$
is an integer), the second and third moments are
\begin{eqnarray}
\NN &=&\N^2\left(1+\bar{\xi}_2\right)\nonumber\\
\NNN&=&\N^3\left(1+3\bar{\xi}_2+\bar{\xi}_3\right),
\label{eq:honint}
\end{eqnarray}
where the volume-averaged connected correlations, $\xibar_2(M)$
and $\xibar_3(M)$, are \citep{berlind_etal03}
\begin{eqnarray}
\bar{\xi}_2 &=&-\frac{N_l(N_l+1)}{\N^2}+\frac{2N_l}{\N}-1\nonumber,\\
\bar{\xi}_3 &=&-\frac{2N_l(N_l^2-1)}{\N^3}+\frac{6N_l^2}{\N^2}-\frac{6N_l}{\N}+2.
\label{eq:xi2xi3}
\end{eqnarray}
For the {\em binomial distribution}
\begin{equation}
P(N=n\vert M)=\frac{{\cal N}_M}{n!({\cal N}_M-n)!}\,p_M^n(1-p_M)^{{\cal N}_M-n},
\label{eq:binomial}
\end{equation} 
with mean $\NM={\cal N}_Mp_M$, the second moment is 
$\NNM={\cal N}_Mp_M({\cal N}_Mp_M-p_M)$ and the higher-order moments are
given by
\begin{equation}
\langle N(N-1)...(N-j)\rangle=\alpha^2(2\alpha^2-1)...(j\alpha^2-j+1)\N^{j+1},
\label{eq:hobinomial}
\end{equation}
where the parameter $\alpha$ is defined as 
\begin{equation}
\alpha^2_M\equiv \NNM/\NM^2.
\label{eq:alpha}
\end{equation}
The function $\alpha^2_M$ is a convenient measure of how different
$P(N\vert M)$ is from the Poisson distribution, for which $\alpha^2_M=1$.
For distributions narrower than the Poisson $\alpha^2_M<1$, while 
for broader distributions $\alpha^2_M>1$.  Semi-analytic models and
hydrodynamic simulations predict a significantly sub-Poisson $\PNM$
distribution at low $\N$ (\citealt{berlind_etal03} and references therein).
Moreover, it has been shown that a sub-Poisson $\PNM$ distribution is 
required in order to produce a correlation function of the observed power-law form.
\citep{benson_etal00a,seljak00,peacock_smith00,scoccimarro_etal01,berlind_weinberg02}

\section{Results}

\begin{figure}[t]
\centerline{\epsfxsize=4truein\epsffile{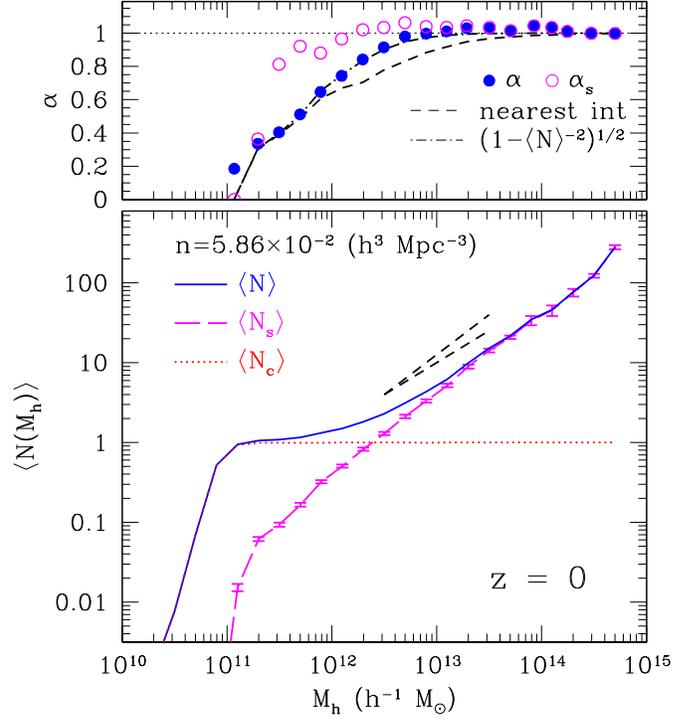}}
\caption{{\it Bottom panel:\/} the first moment of the halo occupation 
  distribution, as a function of host mass for the halo sample with
  number density $n=5.86\times 10^{-2}h^3\rm Mpc^{-3}$ in the
  $\Lambda$CDM$_{80}$ simulation at $z=0$. The {\it solid line} shows
  the mean total number of halos including the hosts, while the {\it
    long-dashed\/} line shows the mean number of satellite halos. The
  error bars show the uncertainty in the mean. The {\it dotted line} shows
  the step function corresponding to the mean number of ``central''
  halos.  Note that by definition, the solid line is the sum of the
  dotted and long-dashed lines. The two short-dashed lines indicate the
  dependencies $\propto M_{\rm h}$ and $M_{\rm h}^{0.8}$. {\it Upper
    panel:} the parameter $\alpha\equiv \NN^{1/2}/\N$ for the full HOD
  ({\it solid points}) and the HOD of satellite halos ({\it open
    points}). The dotted line at $\alpha=1$ shows the case of a Poisson 
  distribution.  Note that the HOD becomes sub-Poisson at small host 
  masses. However, the HOD of satellites remains close to Poisson down to 
  masses an order of magnitude smaller than for the full HOD. Indeed, if the
  satellite HOD is Poisson, $\alpha=(1-1/\N^2)^{1/2}$ for the full HOD
  [see eq.~(\ref{eq:alphahod})]. This expression is shown by the {\it
    dot-dashed} line, which describes the points very well. The full
  HOD at small $M_{\rm h}$ is also well described by the nearest
  integer distribution [see eqs.~(\ref{eq:honint}) and~(\ref{eq:xi2xi3})]
  shown by the {\it dashed line}.
\label{fig:hodfull}}
\end{figure}

\subsection{The Halo Occupation Distribution}
\label{sec:hod}

We start discussion of our results with the factorial moments of the
HOD defined in the previous section. Figure~\ref{fig:hodfull} shows
the first moment of the HOD for the halo sample with number density of 
$5.86\times 10^{-2}h^3\ \rm Mpc^{-3}$ ($V_{\rm max}>70\ \rm
km\,s^{-1}$). Given that the halo samples are constructed by
simply selecting all halos with circular velocities larger than a
threshold value, the HOD will have a trivial component corresponding
to the host halo:
\begin{equation}
N_{\rm c}=\left\{ 
\begin{array}{ll}
1 & \mbox{for $M_{\rm h}\geq M_{\rm min}$}\\
0 & \mbox{for $M_{\rm h}<M_{\rm min}$}
\end{array}
\right.
\label{eq:nh}
\end{equation}
where $M_{\rm min}$ is the mass corresponding to the threshold of the
maximum circular velocity of the sample. The first moment of this
component is simply a step-like function shown in the bottom panel of
Figure~\ref{fig:hodfull} by the dotted line. Note that halo samples
are defined using a threshold $V_{\rm max}$, while the HOD is plotted
as a function of halo mass.  The transition from zero to unity is
therefore smooth because certain scatter exists between $V_{\rm max}$
and halo mass \citep{bullock_etal01b}. We find that the scatter is 
approximately gaussian and its effect on $\langle N_{\rm c}\rangle$ can
be described as
\begin{equation}
\langle N_{\rm c}\rangle = {\rm erf}(5\,[1-M/M_{\rm min}]).
\label{eq:ncerf}
\end{equation}

The second HOD component corresponds to the probability for a halo of
mass $M$ to host a given number of subhalos $N_s=N-1$: $P_s(N_s\vert
M)\equiv P(N_s+1\vert M)$. The first moment of this component is shown
by the long-dashed line.  As we noted above, this separation is
equivalent to differentiating between central and satellite galaxies
in observations or in semi-analytic models.

The first three 
moments of $P_s(N_s\vert M)$
are related to the moments of the overall HOD as follows
\begin{eqnarray}
\Ns&=&\N-1;\label{eq:nsn}\\
\NsNs&=&\NN - 2\N +2;\label{eq:ns2n2}\\
\NsNsNs&=&\NNN-\nonumber\\
&&  3\NN+ 6\left(\N-1\right)\label{eq:ns3n3}.
\end{eqnarray}
As can be seen from Figure~\ref{fig:hodfull}, $\Ns$ has a simple power
law form, while the shape of the full $\N$ (shown by the solid line)
is complicated and consists
of a step, a shoulder, and the high-mass power-law tail. 
The parameter
$\alpha$ plotted in the upper panel indicates that both $P(N\vert M)$
and $P_s(N_s\vert M)$ are close to Poisson at high masses and become
sub-Poisson as the host mass approaches the minimum mass of the
sample. However, the satellite HOD can be described by the Poisson
distribution down to host masses an order of magnitude smaller than
the full HOD.  The latter is well described by the nearest integer
distribution (see eqs.~[\ref{eq:honint}] and~[\ref{eq:xi2xi3}]) at
small $M_{\rm h}$.  This result suggests a simple model for the HOD:
every host halo contains one halo (itself) and a number of satellite
subhalos drawn from a Poisson distribution whose mean is a power-law
function of the host halo mass.

Note that the Poisson shape of the subhalo HOD at small 
masses implies a non-Poisson shape for the overall HOD, as can be seen 
from equations~(\ref{eq:nsn})--(\ref{eq:ns3n3}). For example,
for masses where $P_s(N_s\vert M)$ is Poisson, we have 
\begin{equation}
\alpha^2\equiv \frac{\NN}{\N^2}=1-\frac{1}{\N^2}, 
\label{eq:alphahod}
\end{equation}
which shows that the shape is Poisson ($\alpha^2=1$) at high masses
but drops to zero at low masses.  The HOD thus starts to deviate from
Poisson significantly at $\N\lesssim 4$: e.g, $\alpha^2=0.81$ for
$\N=2.3$, while $\alpha^2=0$ for $\N=1$. As can be seen in the upper
panel of Figure~\ref{fig:hodfull}, equation~(\ref{eq:alphahod})
describes $\alpha$ measured in the simulation very well.

\begin{figure*}[tp]
\vspace{-1cm}
\centerline{\epsfxsize=\textwidth\epsffile{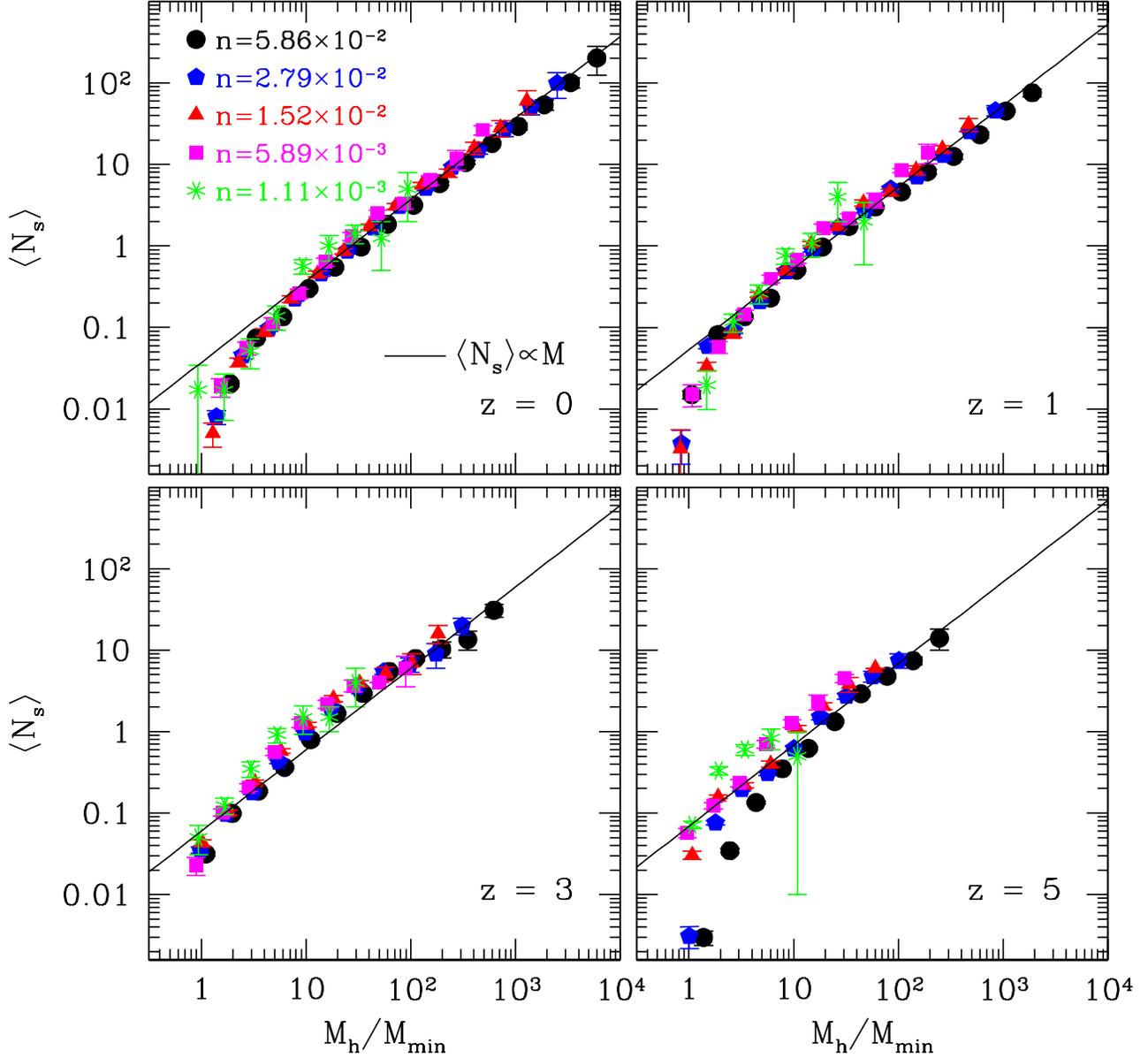}}
\vspace{-1cm}
\caption{The mean number of subhalos
  $\langle N_s\rangle$ as a function of host mass for the halo samples
  of different number densities (symbols of different type) at
  different redshifts. The error bars show the uncertainty in the mean.  The
  mean is plotted as a function of host mass in units of the minimum
  mass of the sample (see Fig.~\ref{fig:vmf}; note that, by definition,
  $\Ns=0$ at $M_{\rm h}/\Mmin=1$ and non-zero points shown at or below
  $M_{\rm h}/\Mmin =1$ are caused by binning).  The mass $M_{\rm h}$
  in the $x$-axis is the mass within the radius corresponding to
  overdensity 180 with respect to the mean density of the universe.
  The number densities in units of $h^3\rm Mpc^{-3}$ are indicated in
  the legend. The solid line in each panel shows the linear relation
  $\langle N_s\rangle\propto M_h$.  The figure shows that the mean number of
  subhalos at different number densities is remarkably similar and
  shows only a mild evolution.  The mass dependence is approximately
  linear $\langle N_s\rangle\propto M_{\rm h}$ for masses $M_{\rm
    h}/M_{\rm min}\gtrsim 5$ (or $\langle N_s\rangle\gtrsim 0.2).$
\label{fig:hodz}}
\end{figure*}

Figure~\ref{fig:hodz} shows the mean of the subhalo HOD, $\langle
N_s\rangle$, for different epochs and samples of different number
densities.  The mean is plotted as a function of mass in units of the
minimum mass of the sample, $\mu\equiv M_{\rm h}/M_{\rm min}$.  The
figure shows that the mean number of subhalos as a function of $\mu$
at different number densities is remarkably similar and exhibits only
a weak evolution with time.  The mass dependence is approximately
linear $\langle N_s\rangle_{\mu}\propto \mu$ for masses $\mu\gtrsim
5$ (or $\langle N_s\rangle\gtrsim 0.2).$ The formal linear fits to
the $\langle N_s\rangle_M$ of individual samples result in best
fit slopes close to unity for $M_{\rm h}/M_{\rm min}\gtrsim 5$ with
rather small deviations from the linear behavior. The best fit slopes
for the samples of different number densities at $z=0$ are $0.99\pm
0.01$, $0.92\pm 0.03$, $0.96\pm 0.08$, $1.04\pm 0.08$, and $0.61\pm
0.21$ in the order of the decreasing number density.  As a function of
redshift, the best fit slopes for the sample of $n=5.86\times
10^{-2}h^3\rm\ Mpc^{-3}$ are $1.03\pm 0.01$, $1.05\pm 0.02$, $1.18\pm
0.05$, $1.28\pm 0.04$, for $z=0,1,3,5$, respectively.

Although $\langle N_s\rangle$ is close to the power law for most of
the most of the mass range, the deviations from power law are evident
at $M\sim M_{\rm min}$, especially at low redshifts. A more accurate
formula describing the first moment is
\begin{equation}
\langle N_s\rangle=(M/M_1 - C)^{\beta},
\label{eq:nsfit}
\end{equation}
where $M_1$ is normalization, defined as $\langle N_s(M_1)\rangle=1$,
and $C$ is a constant for a given redshift. Parameters $M_1$ and $C$
exhibit mild evolution with redshift. For $z=0$, $C\approx 0.045$ or
$M_1/M_{\rm min}\approx 22$.

The slope of the high-mass tail of $\NM$, $\beta$, is one of the key 
factors determining galaxy clustering statistics.  In our results
the asymptotic slope of $\Ns$ in the total mean $\N$ is reached only 
at relatively high masses. A linear fit at intermediate
masses is likely to result in a shallower slope. Thus, for example, a
power-law fit $\N\propto M^{\beta}$ to the full HOD shown in
Figure~\ref{fig:hodfull} for $\N>4$ gives $\beta=0.87\pm 0.01$, while
the fit to $\Ns$ for the same range of masses gives $\beta_s=1.03\pm
0.01$.  This may explain the smaller than unity slopes of the mean
occupation number obtained in several theoretical and observational
analyses (e.g., \citealt{berlind_etal03,zehavi_etal03}). These estimates 
may therefore be underestimates of the true asymptotic high-mass slope. 

\begin{figure}[t]
\vspace{-1cm}
\centerline{\epsfysize4.truein \epsffile{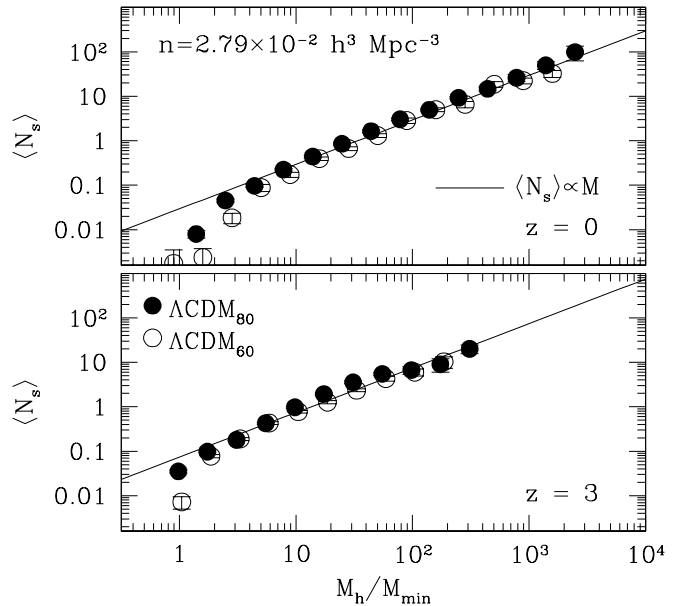}}
\vspace{-0.75cm}
\caption{The mean number of subhalos
  $\langle N_s\rangle$ as a function of host mass for the halo samples
  with number density $n=2.79\times 10^{-2}h^3\rm Mpc^{-3}$ in
  the $\Lambda$CDM$_{60}$ ({\it open circles}) and $\Lambda$CDM$_{80}$
  ({\it filled circles}) at $z=0$ ({\it top panel}) and $z=3$ ({\it
    bottom panel}). The error bars show the uncertainty in the mean and are
  smaller than the symbols.  The mean is plotted as a function of host
  mass in units of the minimum mass of the sample (see
  Fig.~\ref{fig:vmf}).  The solid line in each panel shows the linear
  relation $\langle N_s\rangle\propto M_h$.  The figure shows that the
  HOD for a given $M_{\rm h}/M_{\rm min}$ is not sensitive to 
  normalization of the power spectrum.
\label{fig:hodcomp}}
\end{figure}

Figure~\ref{fig:hodcomp} shows a comparison of the $\NsM$ for the
$n=2.79\times 10^{-2}h^3\rm\ Mpc^{-3}$ halo sample in the
$\Lambda$CDM$_{80}$ and $\Lambda$CDM$_{60}$ simulations at $z=0$ and
$z=3$. The subhalo HODs in the two runs agree well over
the most of the mass range at both epochs. The shape of $P_s(N_s\vert \mu)$,
therefore, is not sensitive to the normalization of the power spectrum.  
The systematically lower values of $\Ns$ at low $M_{\rm h}/M_{\rm min}$ 
in the $\Lambda$CDM$_{60}$ simulation are expected. Halos
in this higher-$\sigma_8$ cosmology form earlier and disruption processes
have more time to operate and lower the number of satellites with masses
comparable to that of the host. This difference can also be partly
due to the limited mass resolution of the simulations. 

\begin{figure*}[tp]
\vspace{-6cm}
\centerline{\epsfxsize=\textwidth\epsffile{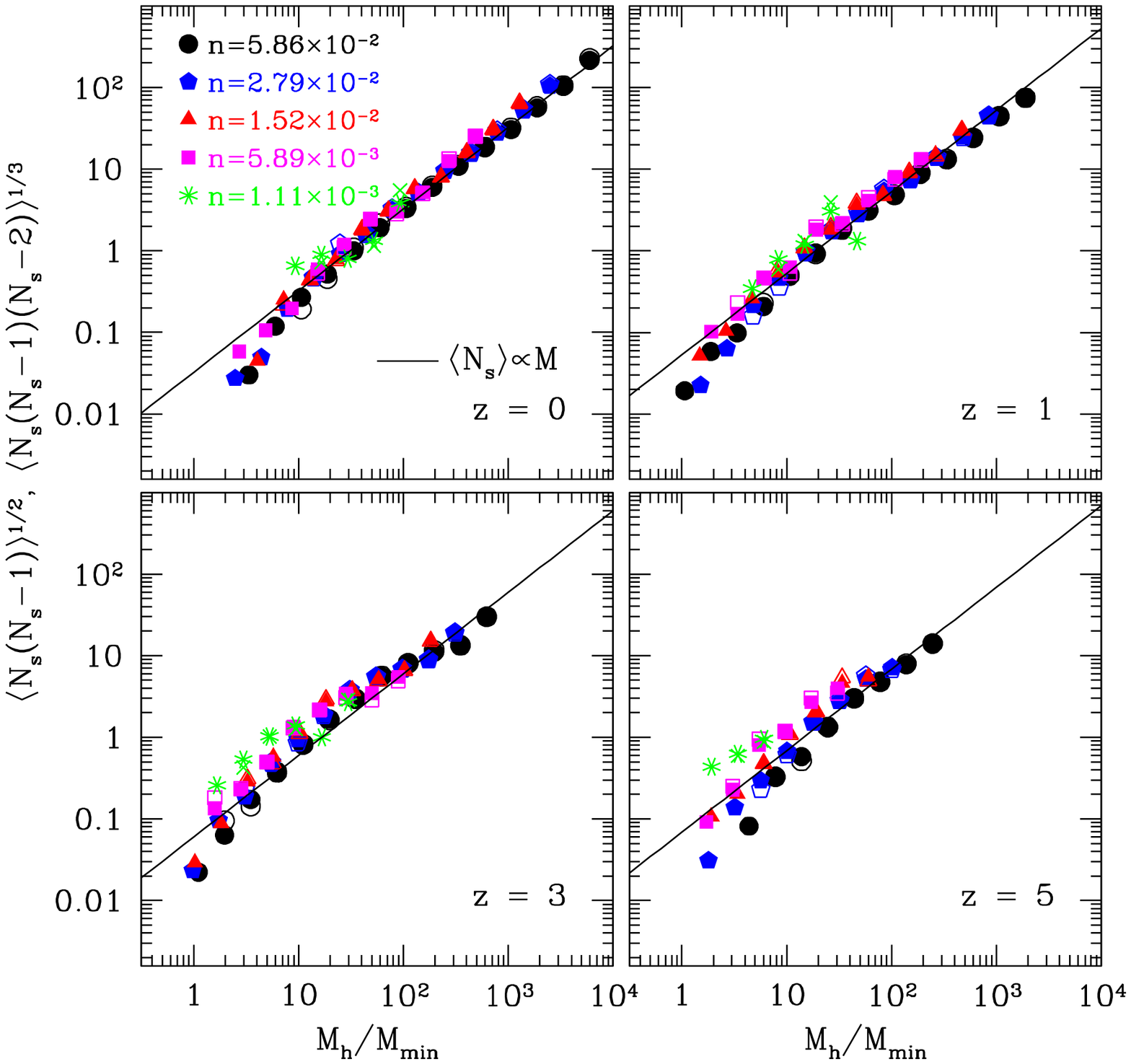}}
\vspace{-1cm}
\caption{The square root of the second moment, 
  $\langle N_s(N_s-1)\rangle^{1/2}$ ({\it filled symbols}), and the
  cube root of the third moment$\langle
  N_s(N_s-1)(N_s-2)\rangle^{1/3}$ ({\it open symbols}, except for the
  lowest number density which is shown by {\it crosses}) for the halo
  samples of different number densities (symbols of different type) at
  different redshifts.  The moments are plotted as a function of host
  mass in units of the minimum mass of the sample.  The number
  densities in units of $h^3\rm Mpc^{-3}$ are indicated in the legend.
  The solid line in each panel shows the linear relation $\langle N_s\rangle\propto
  M_h$ of the same amplitude as the solid line in the corresponding
  panel of Figure~\ref{fig:hodz}. The figure shows that the HODs at
  different number densities and epochs are remarkably similar and are
  close to the Poisson distribution at $M_{\rm h}/M_{\rm min}\gtrsim
  5$ (or $\langle N_s\rangle\gtrsim 0.2-0.3$).
 \label{fig:hodmz}}
\vspace{-1cm}
\end{figure*}

In Figure~\ref{fig:hodmz} we plot the square root of the second and
the cube root of the third moments of the subhalo HOD for the samples
and epochs shown in Figure~\ref{fig:hodz}.  For comparison the solid
lines show the linear function $\Ns\propto \mu$ of the same amplitude
as in the corresponding panel of Figure~\ref{fig:hodz}.
Figure~\ref{fig:hodmz} shows that $\Ns\approx \NsNs^{1/2}\approx
\NsNsNs^{1/3}$ for $\mu\gtrsim 5$, as expected for the Poisson
distribution (eq.~[\ref{eq:hopoisson}]). Therefore, $P_s(N_s\vert\mu)$
can be described by the Poisson distribution at these masses.  As in
the case of the mean, the higher moments of the subhalo HOD have
similar shape and amplitude as a function of $\mu$ for different
number densities and redshifts.

\subsection{The halo 2-point correlation function}
\label{sec:cf}

\begin{figure*}[tp]
\centerline{\epsfxsize=\textwidth\epsffile{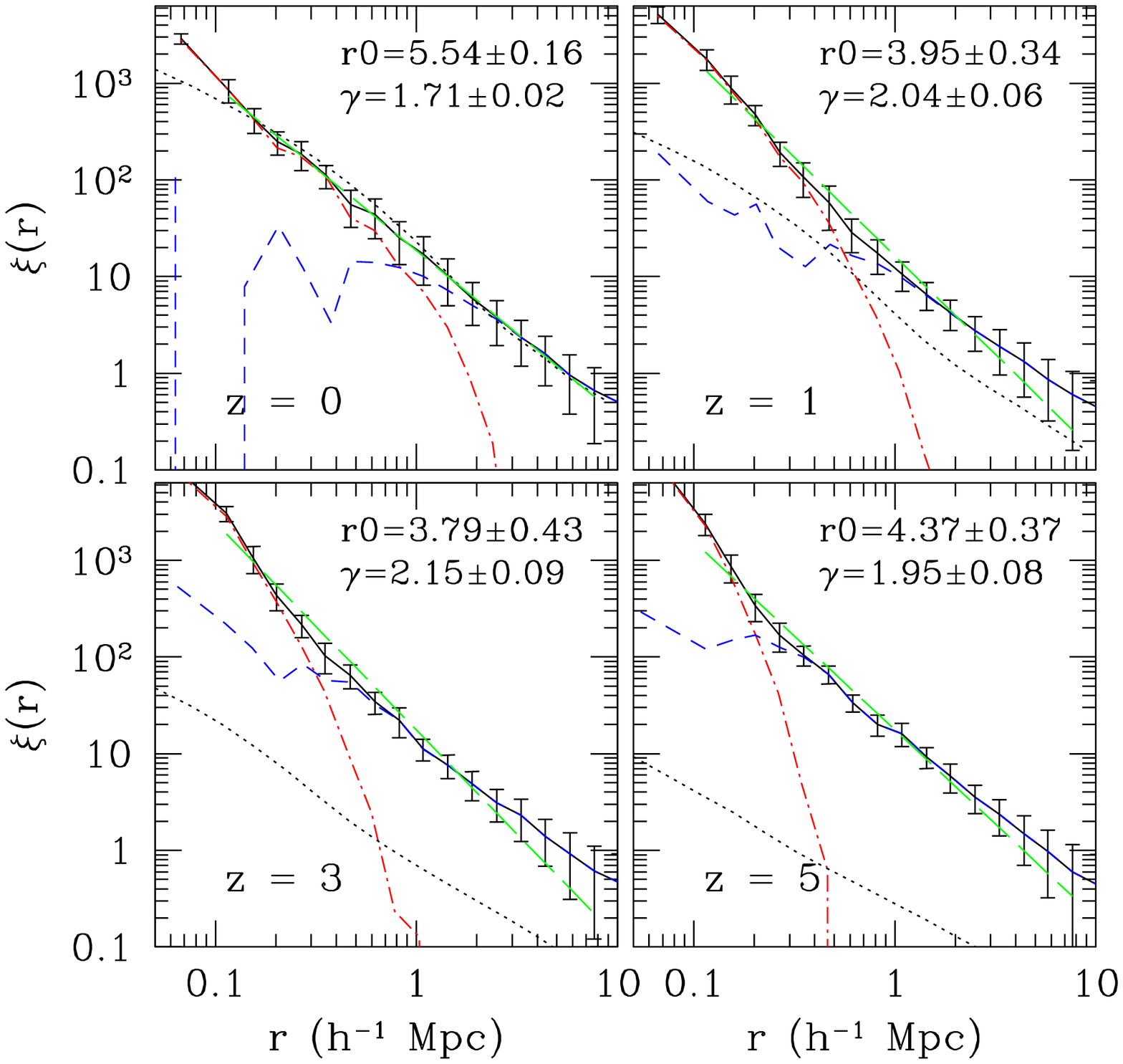}}
\caption{Evolution of the two-point correlation function in the $80h^{-1}\ \rm Mpc$
  simulation. The solid lines with error bars show the clustering of
  halos of fixed number density $n=5.89\times 10^{-3}h^3\ \rm
  Mpc^{-3}$ at each epoch. The error-bars indicate the ``jack-knife''
  one sigma errors, computed using the eight octants of the simulation 
  cube, and are larger than the Poisson error at all
  scales.  The dot-dashed and dashed lines show the corresponding one-
  and two-halo term contributions. The long-dashed lines show the
  power-law fit to the correlation functions in the range of
  $r=\left[0.1-8h^{-1}\ \rm Mpc\right]$. Although the correlation
  functions can be well fit by a power law at $r\gtrsim 0.3h^{-1}\ 
  \rm Mpc$ in each epoch, at $z>0$ the correlation function steepens
  significantly at smaller scales due to the one-halo term.  For
  comparison, the dotted lines show the correlation function of the dark
  matter.
 \label{fig:cfz}}
\end{figure*}

\begin{figure}[tb]
\vspace{-0.5cm}
\centerline{\epsfysize4.25truein\epsffile{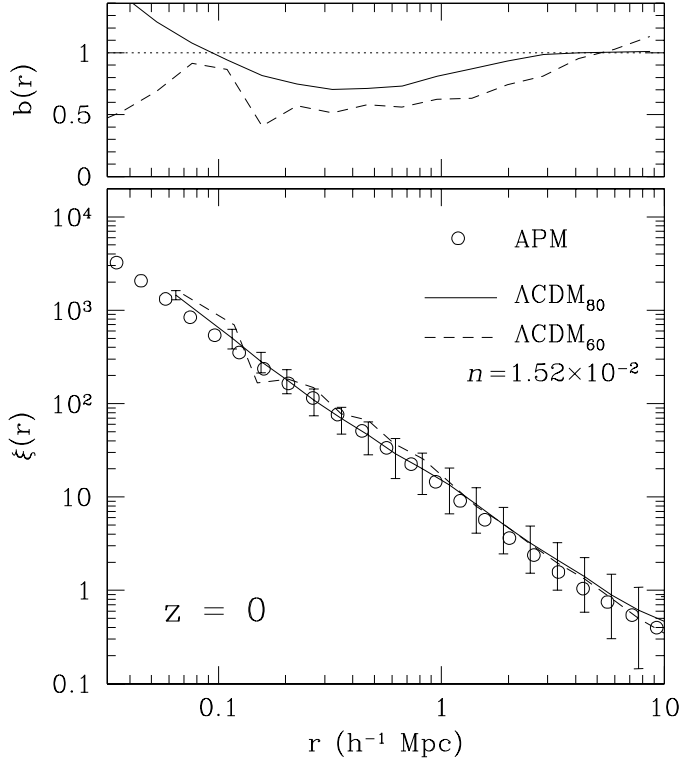}}

\caption{The correlation function and bias for the $n=1.52\times 10^{-2}h^3\ 
  \rm Mpc^{-3}$ sample in the $\Lambda$CDM$_{60}$ ({\it dashed})
  and $\Lambda$CDM$_{80}$ ({\it solid}) simulations. {\it Top
    panel:\/} The bias $b(r)\equiv\sqrt{\xi_{\rm hh}(r)/\xi_{\rm mm}(r)}$.  {\it
    Bottom panel:\/} The halo-halo correlation function in the two
  simulations compared to the correlation function of the APM galaxies
  \citep{baugh96}. The error-bars indicate the ``jack-knife''
  one sigma errors, computed using the eight octants of the simulation 
  cube, and are larger than the Poisson error at all
  scales.
 \label{fig:cfcomp}}
\end{figure}

In this section we present the two-point correlation functions (CFs)
for the halo samples used in the analysis of the HODs.
Figure~\ref{fig:cfz} shows the CFs for the sample of $n=5.89\times
10^{-3}h^3\ \rm Mpc^{-3}$ at $z=5$, $3$, $2$, and $0$, as well as
their one- and two-halo components. The error bars shown for the CFs
are the errors in the mean, estimated from jack-knife resampling using
the eight octants of the simulation cube
\citep[see][]{weinberg_etal02}, and they are dominated by the ``cosmic
variance'' of the finite number of coherent structures in the
simulation volume.  Several interesting features are immediately
apparent. First, at scales $\gtrsim 0.3h^{-1}\ \rm Mpc$ the CFs at all
epochs can be well described by a power law,
$\xi(r)=(r/r_0)^{-\gamma}$, with only mildly evolving amplitude and
slope. The amplitude of the dark matter correlation function, on the
other hand, increases with redshift revealing strongly time-dependent
bias.  At the present-day epoch, there is a slight antibias at
$r\lesssim 1h^{-1}\ \rm Mpc$. Interestingly, the magnitude of the
anti-bias is considerably smaller than in the higher-normalization
($\sigma_8=1$) simulation (see Fig.~7 in \cite{colin_etal99} as well
as Fig.~\ref{fig:cfcomp} in this paper). This is consistent with the
picture where the anti-bias is caused by the halo disruption processes
in high-density regions \citep{kravtsov_klypin99}, as groups and
clusters in the low-normalization model form later and the disruption
processes have less time to operate. The exclusion effect in the
two-halo component is significant at $z=0$, but diminishes at earlier
epochs. This is due mainly to the systematic decrease in the minimum
mass of the sample for the same number density at higher $z$.  The
smaller minimum mass means smaller halo sizes.  The smaller size is
also due to the definition of the virial radius with respect to the
mean density. Even for the same mass higher mean density of the
Universe at higher redshifts results in a smaller virial radius.
Smaller sizes of halos in the sample result in smaller minimum pair
separation for isolated objects. Thus, 2-halo term extends to smaller
$r$.

At $z=0$ the halo CF can be well approximated by a power law at all
probed scales ($0.1-10h^{-1}\ \rm Mpc$). The approximate power-law
shape is due to the relatively smooth transition between the two- and
one-halo components of the CF. At higher redshifts, however, the
transition is more pronounced and occurs at progressively smaller
scales. This results in a significant steepening of the CF at $\sim
0.3-1h^{-1}\ \rm Mpc$. The halo model analysis shows that contribution
of pairs in massive galaxy clusters is critical for a smooth
transition between 1- and 2-halo contributions \citep{berlind_weinberg02}.
At earlier epochs, clusters are rare or non-existent which explains 
a more pronounced transition. 

Indeed, power-law fits using the range of
scales $0.1-8h^{-1}\ \rm Mpc$ give systematically smaller values of
the scale radius $r_0$ and steeper slope $\gamma$ than the fits over
range $\sim 0.3-8h^{-1}\ \rm Mpc$, as can be seen in
Figure~\ref{fig:r0nz13}. All of the fits for the $\Lambda$CDM$_{80}$ 
simulations are performed at $r\leq 8h^{-1}\ \rm Mpc$, as the CF shape
becomes affected at scales larger $>0.1L_{\rm box}$ \citep{colin_etal97,colin_etal99}. We also checked this by comparing matter correlation functions
in the simulation to the model of \citet{smith_etal03}. We find 
that simulation results agree well with the model at scales $<0.1L_{\rm box}$ 
at $z=0$ and at $<0.2-0.3L_{\rm box}$ at higher redshifts.

The power-law shape of the correlation function is rather remarkable,
as it appears to result from a sum of non-power law components.  We
checked the components of the correlation function due to pairs of
different types: central-satellite, satellite-satellite, and
central-central pairs.  The component CFs have a variety of shapes all
deviating strongly from power law.  Yet, the sum is close to the power
law. This indicates that the power-law shape of the galaxy correlation
function may well be a coincidence, as noted by \citet{benson_etal00a} and
\citet{berlind_weinberg02}.

\begin{figure}[tb]
\vspace{-0.5cm}
\centerline{\epsfysize4.truein\epsffile{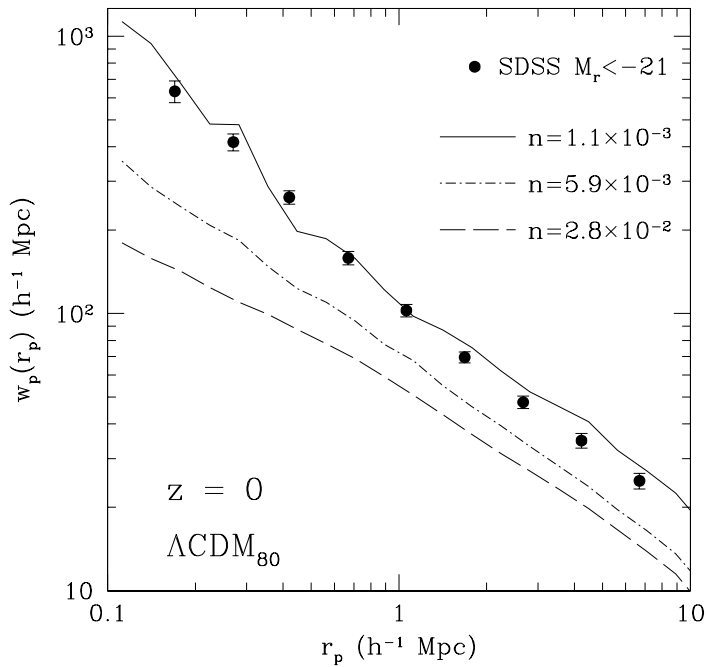}}

\caption{The projected correlation function of the bright ($M_r<-21$) galaxies
  in the SDSS volume-limited sample \citep{zehavi_etal03} compared to
  the $z=0$ projected correlation function of halo samples of three
  different number densities indicated in the legend. Note that
  $M_r<-21$ galaxies have number density of $1.1\times 10^{-3}h^3\ \rm
  Mpc^{-3}$. The figure shows that the correlation functions of
  galaxies and halos of the same number density are in good agreement
(see text for discussion).
 \label{fig:cfs}}
\end{figure}

Comparing the correlation functions in the $\Lambda$CDM$_{60}$ and
$\Lambda$CDM$_{80}$ simulations (Figure~\ref{fig:cfcomp}), we find that
the correlation functions of objects with the same number density are
similar.  This is not surprising in light of the approximate
universality of the HOD demonstrated in the previous section (see
Figs.~\ref{fig:hodz} and \ref{fig:hodcomp}). Figure~\ref{fig:cfcomp}
also shows that the amplitude and shape
of the CF at $z=0$ is in good agreement with that of the galaxies in
the APM survey. As noted by
\citet{kravtsov_klypin99} and \citet{colin_etal99}, the close
agreement of halo and galaxy correlation functions indicates that the
overall clustering of the galaxy population is determined by the
distribution of their dark matter halos.

Figure~\ref{fig:cfs} shows a comparison of the projected
correlation functions:
\begin{equation}
w_p(r_p)=2\int\limits^{r_{\rm max}}_0\xi([r_p^2+y^2]^{1/2})dy,
\label{eq:wp}
\end{equation}
in the volume-limited sample of bright, $M_r<-21$, galaxies
\citep{zehavi_etal03} and halo samples of three representative number
densities in the $\Lambda$CDM$_{80}$ simulation. The upper integration
limit was set to $r_{\rm max}=40h^{-1}\ \rm Mpc$, to mimick the
procedure used to estimated the observed CF. In taking the projection
integral, we extrapolate the simulated CF from $8h^{-1}$~Mpc (the
largest reliable scale of the simulation) to large scales using the
correlation function of dark matter predicted by the
\citet{smith_etal03} model rescaled to match the amplitude of the halo
correlation function at $8h^{-1}$~Mpc.  Note that the SDSS galaxies
have a number density of $1.1\times 10^{-3}h^3\ \rm Mpc^{-3}$ and
their CF should therefore be compared to the solid line. The figure
shows that the correlation functions of galaxies and halos in our
simulations agree remarkably well. In particular, the steepening of
the observed correlation function at $r\lesssim 1h^{-1}\ \rm Mpc$ is
reproduced.  The difference at large scales is not significant given
the large sample variance errors in the halo correlation function that
result from the small size of the simulation cube (see
Figures~\ref{fig:cfz} and \ref{fig:cfcomp}).

\begin{figure}[tb]
\vspace{-0.5cm}
\centerline{\epsfysize4.25truein\epsffile{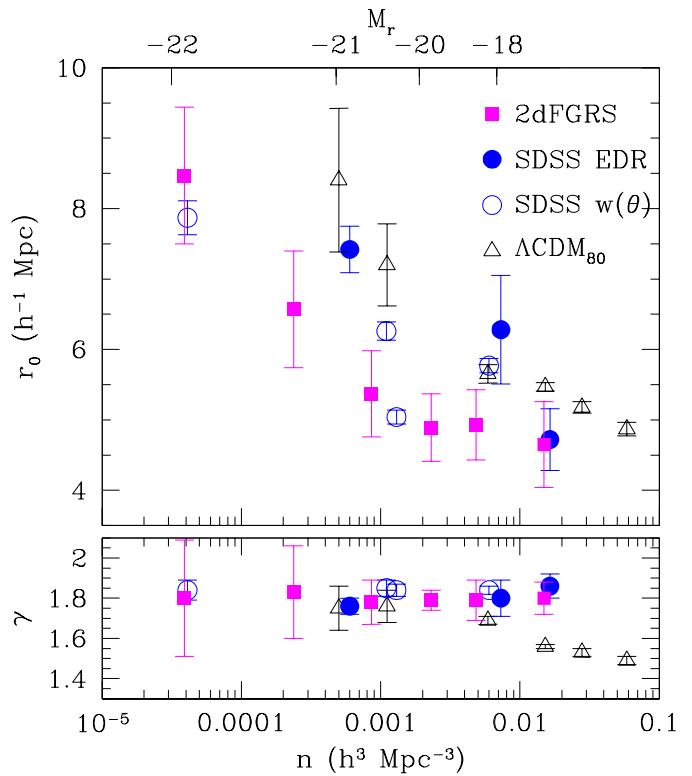}}
\caption{{\it Top panel:\/} the best-fit correlation scale $r_0$ as 
  a function of number
  density at the present day epoch. The results for the dark matter
  halos (open triangles) are compared to the recent measurements of
  galaxy clustering in the SDSS (solid circles are from the analysis
  of the Early Data Release by \citet{zehavi_etal02}, while open
  circles are derived from the analysis of the angular correlations by
  \citet{budavari_etal03}) and 2dF surveys \citep[solid
  squares;][]{norberg_etal02}. The upper axis shows the
  $r$-band absolute magnitude for the SDSS galaxies corresponding to each
  number density. The power-law fits were done over the range of
  scales from $0.3$ to $8h^{-1}\ \rm Mpc$. {\it Bottom panel:\/} the best-fit
  slope of the correlation function as a function of number density.
 \label{fig:r0nz0}}
\end{figure}

Large galaxy redshift surveys have been used to detect a luminosity
dependence of galaxy clustering
\citep[e.g.,][]{guzzo_etal97,willmer_etal98,norberg_etal01,zehavi_etal02}.
There are indications that a similar dependence exists at early epochs
\citep[e.g.,][]{giavalisco_dickinson01}.  As the luminosity of galaxies is
expected to be tightly correlated with the halo maximum circular
velocity or mass, the luminosity-dependence of clustering should be
reflected in the mass-dependence of halo clustering.
Figures~\ref{fig:r0nz0} and \ref{fig:r0nz13} show the best-fit correlation
length $r_0$ and slope $\gamma$ as a function of sample number density, with
lower number densities corresponding to samples of halos with larger
mass (i.e., larger values of threshold $V_{\rm max}$).
Figure~\ref{fig:r0nz0} shows the $z=0$ results compared to the 2dF
\citep{norberg_etal01} and SDSS galaxy surveys
\citep{zehavi_etal02,budavari_etal03}.  The figure
shows that the dependence of halo clustering on sample number density
in our simulation is in general agreement with the SDSS Early Data
Release results \citep{zehavi_etal02}.

The simulation points are systematically higher than the 2dF and
\citet{budavari_etal03} results. Note, however, that the upturn in the
clustering amplitude occurs at approximately the same number density,
$n\approx2-4\times 10^{-3}h^3\ \rm Mpc^{-3}$, in the simulation and
2dF survey.  The difference in amplitude can likely be attributed to
the fact that 2dF galaxies are selected using a blue-band magnitude,
since several recent studies have shown that redder galaxies are
clustered more strongly \citep[e.g.,][]{norberg_etal01,zehavi_etal02}.  
In addition, the halo samples include all objects above a threshold
circular velocity, while most of the observational points in
Figure~\ref{fig:r0nz0} are defined for galaxies in (broad) luminosity
ranges.

Interestingly, all of the observational estimates indicate that the
slope of the CF does not depend strongly on the luminosity. The slope
of the halo CF is in agreement with observations for $n<0.01h^3\ \rm
Mpc^{-3}$ but becomes shallower for smaller mass objects. At $n\approx
0.02h^3\ \rm Mpc^{-3}$ the slope for the halo sample is significantly
shallower than that for the galaxies in the 2dF and SDSS surveys.
This indicates that luminosity dependence of the CF slope may provide
additional useful constraints on galaxy formation. The exercise
demonstrates that both slope and correlation length should be compared
when model predictions are confronted with observations.

Mass dependence of the clustering amplitude is also found at earlier 
epochs (Fig.~\ref{fig:r0nz13}).  The clustering of halos at $z=3$ 
is in reasonable agreement with clustering of Lyman break galaxies 
\citep[LBGs,][]{adelberger00,adelberger_etal03}. The detailed comparison
is complicated due to the often contradictory results from analyses
that use different LBG samples and methods \citep*[for a summary of
recent results see][and \S~\ref{sec:discussion}
below]{bullock_etal02}. An important point is that, as we noted above,
at higher redshifts the steepening of the CF at small scales biases
results if a single power-law is fit down to small scales.  This
bias can be clearly seen in Figure~\ref{fig:r0nz13}, which shows the 
best-fit scale radius and slope for the power-law fits down to both
$0.1h^{-1}\ \rm Mpc$ and $0.3h^{-1}\ \rm Mpc$.  At both $z=1$ and
$z=3$, fits to smaller scales result in smaller $r_0$ and larger
absolute values of the slope $\gamma$.

\section{Discussion}
\label{sec:discussion}

\subsection{Implications for galaxy clustering}

The results presented in the previous sections show that the main
properties of the observed clustering of galaxies are imprinted in the
distribution of their surrounding halos. The term halo here includes
both the isolated halos and the distinct gravitationally-bound
subhalos within the virialized regions of larger systems.

The shape and evolution of the two-point correlation
function and the mass-dependence of clustering amplitude for these
subhalos are in good agreement with constraints, for galaxies of
similar number densities, from recent observational surveys at a range
of redshifts.  In addition, the halo occupation distribution derived
in this study for the subhalos is similar to the HOD obtained for the
galaxies in semi-analytic analyses and for cold gas clumps in
gasdynamics simulations \citep{berlind_etal03}.  All of these are
consistent with present observational constraints on the galaxy HOD
\citep[e.g.,][]{zehavi_etal03}, for galaxies of similar number
densities; however, observational measurements of the HOD are not
yet robust enough to make a meaningful comparison.  The test of our 
predictions against real data must thus await future measurements.

This result has several implications. First and foremost, it means that
the formation of halos and their subsequent merging and dynamical
evolution are the main processes shaping galaxy clustering. This
appears to be true for the clustering of halo samples with maximum
circular velocities larger than a given threshold value, which were
the focus of the present and other recent studies
\citep{kravtsov_klypin99,colin_etal99,neyrinck_etal03}. This type
of halo sample should
correspond to volume-limited samples of galaxies with luminosities
above a certain threshold value because the maximum circular velocity
of a halo is expected to be tightly correlated with the luminosity of
the galaxy it hosts. The caveat, of course, is that 
we can expect a considerable band-dependent scatter 
between galaxy luminosity and $V_{\rm max}$, which needs to be accounted for
in the model. The inclusion of scatter is relatively straightforward and
future analysis will show just how large the effect of scatter is.

Useful constraints on galaxy evolution and better
understanding of galaxy clustering can be obtained by more
sophisticated analyses. For example, it would be interesting to
compare the HOD of galaxies of different colors
\citep[e.g.,][]{zehavi_etal02} with that of halos with different
merger histories or environments, which would likely provide insight into
the formation of galaxies of different types. 

\begin{figure*}[tp]
\centerline{\epsfxsize=4truein\epsffile{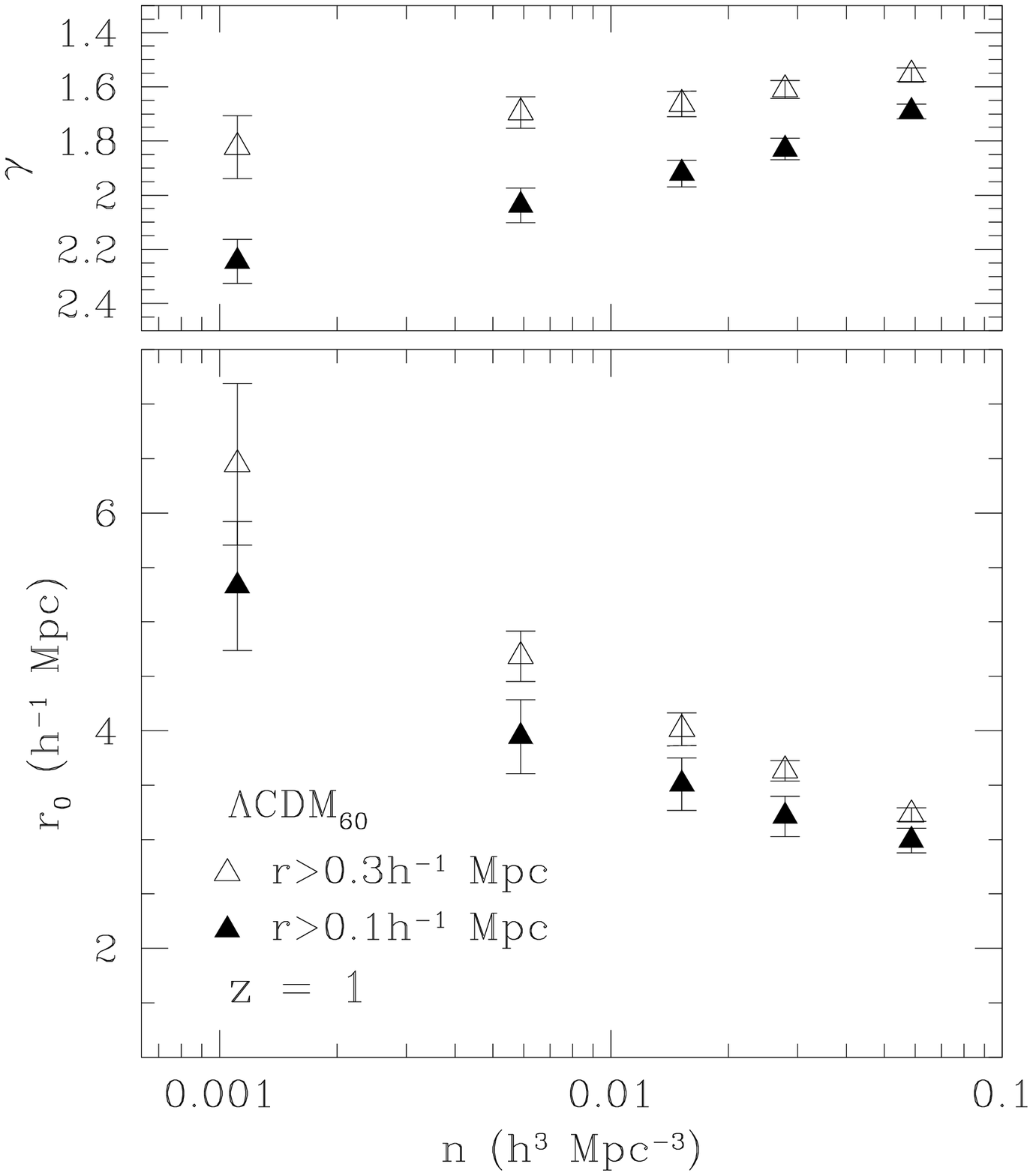}
\hspace{-1cm}
\epsfxsize4truein\epsffile{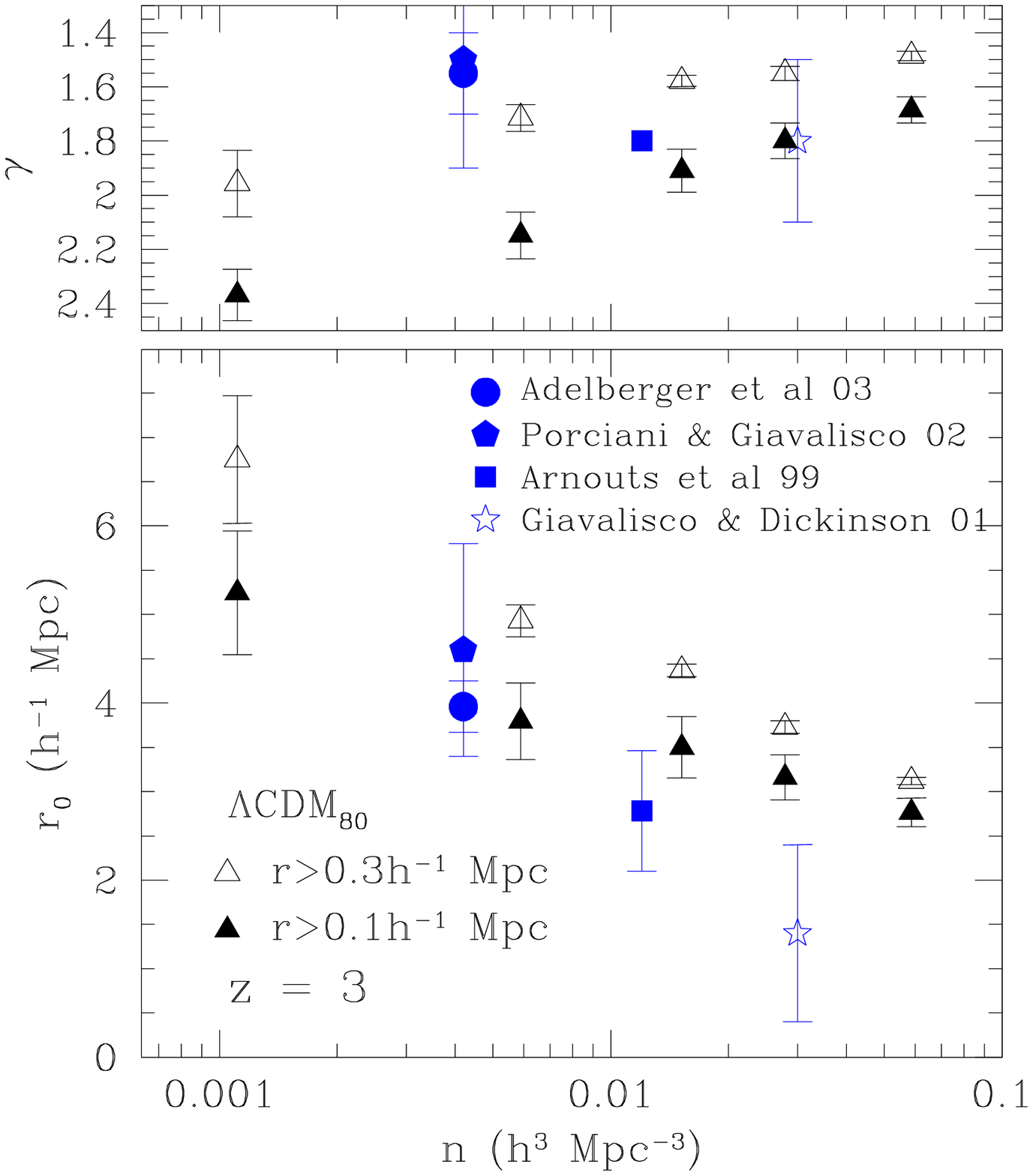}}
\caption{The same as in Fig.~\ref{fig:r0nz0} but for $z=1$ and $z=3$. 
  At high redshifts the correlation function significantly steepens at
  small separations due to the one-halo correlations (see
  Fig.~\ref{fig:cfz}).  This makes the results of the power-law fit
  sensitive to the minimum separation $r_{\rm min}$ used in the fit.
  The figure shows results for two values $r_{\rm min}$: $100h^{-1}\ 
  \rm kpc$ (solid triangles) and $300h^{-1}\ \rm kpc$ (open
  triangles). At $z=3$ the results are compared to the recent observational 
  measurements of Lyman Break Galaxy clustering at $z\sim 3$.
 \label{fig:r0nz13}}
\end{figure*}

One of the most interesting features of galaxy clustering is the
approximately power-law shape of the two-point correlation function.
Although departures from a power law have been found
\citep{zehavi_etal03}, they are quite small. This has been and still
is a major puzzle in galaxy formation studies, especially in light of
the strong deviations from the power-law behavior seen in the dark
matter correlation function \citep{klypin_etal96,jenkins_etal98}.  In
the framework of the recently developed halo model, there also does
not seem to be a generic way to produce a purely power-law CF
\citep{berlind_weinberg02}. The power-law shape thus appears to be
somewhat of a coincidence.

Without a doubt, the fact that the (approximately) power-law CF observed
for galaxies is reproduced with subhalo populations identified in
simulations with the correct slope and amplitude at $z=0$ down to
scales $<100h^{-1}\ \rm kpc$ can be viewed as a significant success.
The power-law nature of the correlation fucntion is due to a relatively 
smooth transition between its one- and
two-halo terms.  At higher redshifts this transition is more
pronounced and the CF steepens at small scales.  The transition scale
decreases with increasing redshift, reflecting the decrease in the
average halo size with time.  The key to the puzzle of the power-law
shape of the CF thus appears to be in understanding the transition
between the one- and two-halo terms.

Interestingly, if similar large departures from a power law exist
for real high-redshift galaxies, this may bias observational power-law
fits and explain some of the discrepancies between observational
analyses.  For example, as can be seen in Figure
\ref{fig:r0nz13}, single power-law fits to smaller radii result in
systematically smaller values of $r_0$ and steeper slopes $\gamma$. If
the slope is kept fixed, as is often done in analyses of high-$z$
galaxy clustering, the derived correlation length may be artificially large.
Using the halo model formalism, \citet{zheng03} recently showed that the
large correlation length inferred for red galaxies at $z\approx 3$ by
\citet{daddi_etal03} can be explained by the steepening of the CF at
small scales, as observed for subhalos in our simulation. \citet{daddi_etal03}
studied clustering of red galaxies in the Hubble Deep Field South. The
number density and range of radii probed in their study are $n\approx
7\times 10^{-3}h^3\ \rm Mpc^{-3}$ and $\sim 0.04-1h^{-1}$, with the
statistically significant correlations detected only for $r\lesssim
400h^{-1}\ \rm kpc$. All scales here and below are computed assuming the flat
$\Lambda$CDM cosmology adopted in this paper.  Given a limited number
of radial bins, \citet{daddi_etal03} used a power-law fit to the
angular correlation function with a fixed slope of $-0.8$ and obtained
the best fit correlation length of $r_0\approx 8h^{-1}\ \rm Mpc$.

As a comparison, for the halo sample with the number density of
$6\times 10^{-3}h^3\ \rm Mpc^{-3}$ in our simulation, a weighted least
squares fit over the interval $40-400h^{-1}\ \rm kpc$ with the slope
fixed to $-1.8$ gives $r_0=7.85\pm 0.77h^{-1}\ \rm Mpc$,
remarkably close to the value derived by \citet{daddi_etal03}.  This,
however, is simply a reflection of steepening of the CF at small
scales. If both the slope and the correlation length are allowed to vary, best
fit values for the same range of scales are $r_0 =2.04\pm 0.76h^{-1}\ 
\rm Mpc$ and $\gamma=2.69\pm 0.18$. At the same time, the correlation
function is unity at $r_0\approx 5h^{-1}\ \rm Mpc$ and the power-law
fit over the interval $0.3-8h^{-1}\ \rm Mpc$ gives $r_0= 4.93\pm
0.49h^{-1}\ \rm Mpc$ and $\gamma = 1.72\pm 0.13$ (see
Fig.~\ref{fig:r0nz13}), in very good agreement 
with the halo model calculations of \citet{zheng03}. 

Similar biases may explain some of the discrepant results on the
clustering of Lyman break galaxies (LBGs), virtually
all of which have assumed a power-law correlation function, and many 
of which have fixed the power-law slope during fits. Many studies
presented seemingly contradictory measurements of correlation lengths
ranging from 1 to $5h^{-1}$~Mpc 
\citep{giavalisco_dickinson01, bullock_etal02,adelberger_etal03,
porciani_giavalisco02,arnouts_etal99}.
The variety of values and discrepancies of the correlation length and slope 
could be explained by departures of the high-$z$ CF
from the single power law.  Thus, for example, the power-law fit over the
smallest angular scales in the HDF sample by
\citet{giavalisco_dickinson01} gives the smallest value of $r_0$ and
the steepest slope $\gamma$. The analysis of \citet{arnouts_etal99}
fits the CF over a larger range of scales with the slope fixed at a
relatively low value and results in a larger value of $r_0$. This would
be expected if the CF steepens at small scales, as observed in our
simulations.  All of the studies perform the Limber deprojection
assuming a power-law CF, which may further bias results. The
implication is that if the deviations from the power law for the
galaxy CF are as strong as indicated by our results, the assumption of
the single power law is dangerous and is likely to bias results. The
magnitude of the bias depends on the range of scales probed and the
analysis method.  

On the positive side, the sharper transition from one- to two-halo
components of the CF at early epochs means that departures from the
power law may be easier to detect in high-redshift surveys than 
they are at $z=0$. Such features can be useful in understanding the
environments and nature of galaxies because the one-halo term contains
information about $\PNM$ and the radial distribution of galaxies in
halos. For example, we can expect the distribution of red galaxies to
be biased toward the high-density regions of groups and clusters.
Their correlation function is therefore expected to have a more
pronounced one-halo term. Although the sizes of samples are still
relatively small at present, it is interesting that the {\it
  projected} CF of the largest LBG galaxy samples to date presented by
\citet{adelberger_etal03} \citep[their Fig.~23; see
also][]{hamana_etal03} indicates a departure from
the power law similar to that in $z=3$ panel of Fig.~\ref{fig:cfz}.

\subsection{The shape of the halo occupation distribution}

The main result of our study is the approximate universality of the
halo occupation distribution, $P_s(N_s\vert \mu)$ where $\mu\equiv
M/M_{\rm min}$, for the subhalos in our samples. We show that the
overall HOD can be split into the probability for a halo of mass $M$
to host a central galaxy and its probability to host a given number
$N_s$ of satellite galaxies, which significantly simplifies the halo
model.  The former can be approximated by a step function, while the
latter is well approximated by the Poisson distribution fully
specified by its first moment $\Ns$ for $\mu\gtrsim 5$.  The first
moment of the distribution is well represented by a simple power law
$\N\propto \mu^{\beta}$ for $\mu\gtrsim 5$ with $\beta$ close to unity
for a wide range of number densities and redshifts.  We also find that
the form of the satellite HOD is not sensitive to the normalization of
the power spectrum. Note that although the amplitude of $\N(\mu)$
changes little for different number densities, redshifts, and spectrum
normalizations, some weak dependencies do exist. Thus, for example,
the first moment for the lower number density (i.e., higher mass)
samples have a systematically higher amplitude. Moreover, there is a 
factor of three increase in the normalization from $z=0$ to $z=5$. The 
HOD moment in the lower-normalization simulation has a slightly higher
amplitude than in the higher normalization run
(Fig.~\ref{fig:hodcomp}).

It is worth noting that results presented here for the dark
matter halos are in good agreement with the results of 
semi-analytic and gasdynamics simulations obtained by
\citet{berlind_etal03}. In particular, we checked that our results on the HOD of the
satellite galaxies are in very good agreement with the results of
these simulations.

The approximately linear dependence of the first moment on the host
mass, $\Ns\propto M$, is related to the shape of the subhalo mass and
velocity functions: $N_s(>V_{\rm sub})=A({V_{\rm sub}/V_{\rm
    h})^{-\eta}}$, where $N_s$ is the number of subhalos within the
virial radius of the host and $V_{\rm sub}$ and $V_{\rm h}$ are the maximum
circular velocity of subhalos and the host, respectively.
High-resolution simulations give $\eta\approx 3$
\citep{klypin_etal99,moore_etal99} and normalization approximately
independent of mass $A\approx {\rm const}$
\citep{moore_etal99,colin_etal03}. Therefore, the number of subhalos with
$V_{\rm sub}$ above a certain threshold $V_{\rm th}$ scales as
$N_s(>V_{\rm th}\vert M_h)=A V_{\rm th}^{-\eta} V_{\rm h}^{\eta}$.  The
circular velocity of isolated host halos is tightly correlated with
their mass $M_{\rm h}=C V_{\rm h}^a$ with $a\approx 3-3.3$
\citep[e.g.,][]{avila_reese_etal99,bullock_etal01b}. We thus have
$N(>V_{\rm th}\vert M_h)\propto M_{\rm h}^{\beta}$ with
$\beta={\eta/a}\sim 1$.

The simple combination of a step function representing the central
galaxies and a Poisson distribution for the satellite galaxies can be
compared to the sum of the HODs for the two, which is in general
significantly more complicated, consisting of a step, a shoulder, and
a power-law high mass tail, as observed in our simulations and in
simulations studied by \citet{berlind_etal03}.  This shape is
sometimes approximated simply by a power law
\citep[e.g.,][]{bullock_etal02} or by the combination of a step
function and a power law for masses larger than some $M_p$
\citep[e.g.,][]{zehavi_etal03}.  However, these are only crude
approximations, as the first moment of the HOD in simulations is almost
never flat, especially at high redshifts.  Modeling of the HOD as a
combination of host and satellite HODs provides a considerably more
accurate prescription without increasing the number of parameters. It
is also more physically motivated because it is reasonable to expect
that the processes that control the formation of the central galaxy
are different from those that control the abundance of satellite
galaxies. 

Most importantly, the simple form of the satellite HOD hints at some
simple physical processes that control the satellite population.
Understanding these processes is well within the capabilities of
current numerical simulations or semi-analytic models for satellite
accretion and orbital evolution
\citep[e.g.,][]{bullock_etal00,zentner_bullock03}. Further theoretical
studies of the satellite HOD should thus provide key clues to the
understanding of small-scale galaxy bias and the power-law shape of
the correlation function.

\subsection{Halo occupation distribution model}
\label{sec:hodmodel}

Our results suggest a simple yet accurate 
model for the halo occupation distribution for samples
selected with a given mass or luminosity threshold, which could be used 
in theoretical modeling and model fits to the observational data. 
The probability for a halo of mass $M$ to host $N$ galaxies, $P(N\vert M)$,
is split into probability to host one central galaxy $P_{\rm c}(M)$
and $N_s$ satellite galaxies $P_s(N_s\vert M)\equiv P(N_s+1\vert M)$. 
In the simplest case when halos are selected using a
quantity tightly correlated with mass (e.g., the maximum circular
velocity in our analysis), the distribution $P_{\rm c}(M)$ can be 
approximated by a step function (eq.~[\ref{eq:nh}]) changing from zero 
to unity at $M_{\rm min}$, the mass corresponding to the threshold
quantity.  If selection is done using a quantity correlated with mass
with a significant scatter, such as galaxy luminosity, the transition
from zero to unity at $M\gtrsim M_{\rm min}$ will be smoother.  The
transition can be modeled to take into account the scatter in the
mass-luminosity relation and other sample selection effects. For example, 
equation~(\ref{eq:ncerf}) above shows how a gaussian scatter between 
luminosity and mass could be taken into account. 

The probability distribution $P_s(N_s\vert M)$ is Poisson for
$M>M_{\rm min}$ and is defined by its first moment, given by
\begin{equation}
\Ns = (M/M_1)^{\beta},
\end{equation}
and by higher moments, given by eq.~(\ref{eq:hopoisson}).
Our results (Figs.~\ref{fig:hodz} and \ref{fig:hodmz}) give $M_1\approx 30M_{\rm min}$
for $z=0$, 
$20M_{\rm min}$ for $z=1$, $10M_{\rm min}$ for $z=3,5$ and $\beta\approx 1$ for all number densities and redshifts. A more accurate
formula describing the first moment is given by equation~(\ref{eq:nsfit}).

To relate this HOD model to a population of galaxies with a known
spatial number density $\bar{n}$ in a given cosmology we have:
\begin{eqnarray}
\bar{n}(>M_{\rm min}) &=& \int_{M_{\rm min}}^\infty \N(M,M_{\rm min})n_{\rm h}(M)dM\nonumber\\
          &\approx & \int_{M_{\rm min}}^\infty \left[1 + (M/M_1)^{\beta}\right]n_{\rm h}(M)dM,
\label{eq:nhalo}
\end{eqnarray}
where $n_{\rm h}(M)$ is the theoretical mass function of host halos
\citep[e.g.,][]{sheth_tormen99}. This function can be inverted to
estimate $M_{\rm min}$.  An approximate estimate of $M_{\rm min}$ can
be obtained by using the approximation to the subhalo$+$host mass
function: $n(>M_{\rm min})\approx n_{\rm h}(>M)(1-f_{\rm sub})^{-1}$,
instead of the integral in eq.~(\ref{eq:nhalo}). Here, $f_{\rm
  sub}\approx 0.15-0.25$ for masses in galactic range (see
Fig.~\ref{fig:vmf}).

The first moment of the HOD distribution is $\N=\Ns+1$,
while the second and third moments and higher moments can be specified 
completely in terms of moments of $P_s(N_s\vert M)$ as 
\begin{eqnarray}
\NN&=&\NsNs+2\Ns\nonumber\\
   &=&\Ns(\Ns+2),\nonumber\\
\NNN&=&\Ns^2(\Ns+3), {\rm etc.}
\end{eqnarray}
The HOD moments can be used to calculate clustering statistics in the 
framework of the halo model. 

The model has at most three free parameters: the minimum mass $M_{\rm min}$
of a halo that can host a galaxy in the sample, and the normalization
and slope of the first moment of the satellite galaxy HOD
$(M/M_1)^{\beta}$, where $M_1$ is the halo mass corresponding to the
average of one satellite galaxy. The model provides more accurate 
description of the HOD in simulations and semi-analytic models
compared to other models used in the literature, without an increase
in the number of free parameters. 

\section{Conclusions}
\label{sec:conclusions}

In this study we analyze the halo occupation distribution (HOD)
and two-point correlation function (CF) of dark matter halos 
using high-resolution dissipationless simulations of the
concordance $\Lambda$CDM model. Our main conclusions can 
be summarized as follows. 

\begin{itemize}
\item We find that the shape of the HOD, the probability distribution
  for a halo of mass $M$ to host a number of subhalos $N$, $P(N\vert
  M)$, is similar to that found for galaxies in semi-analytic and 
  $N$-body$+$ gasdynamics studies \citep{berlind_etal03}. 
  
\item The first moment of the HOD, $\NM$, has a complicated shape
  consisting of a step, a shoulder, and a power-law high mass tail.
  The HOD can be described by Poisson statistics at high halo masses
  but becomes sub-Poisson for $\NM\lesssim 4$.  We show, however, that
  this behavior can be easily understood if the overall HOD is thought
  of as a combination of the probability for a halo of mass $M$ to
  host a central galaxy, $P_{\rm h}(M)$, and the probability to host a
  given number $N_s$ of satellite galaxies, $P_s(N_s\vert M)$.  The
  former can be approximated by a step-like function, with $\Nc=1$ for
  $M_h>\Mmin$, while the latter can be well approximated by the
  Poisson distribution, fully specified by its first moment. The first
  moment can be well described by a simple power law $\Ns\propto
  \mu^{\beta}$ for $\mu\gtrsim 5$ ($\mu\equiv M/M_{\rm min}$) with
  $\beta$ close to unity for a wide range of number densities and
  redshifts. 
  
\item We find that the satellite HOD, $P_s(N_s\vert \mu)$, has a
  similar amplitude and shape for a wide range of halo number
  densities and redshifts. It is also not sensitive to the
  normalization of the power spectrum for objects of a fixed number density.
  
\item We study the two-point correlation function of galactic halos at
  scales $0.05-8h^{-1}\ \rm Mpc$. We confirm and extend results of our
  previous studies \citep{kravtsov_klypin99,colin_etal99} based on
  lower resolution simulations, in which we found that 1) the halo
  correlation function can be well described by a power law at
  scales $\gtrsim 300h^{-1}\ \rm kpc$ at all epochs; 2) the amplitude
  of the correlation function evolves only weakly with time; and 3) the
  evolution results in a small-scale anti-bias at the present day
  epoch. 
  
\item We find that the small-scale anti-bias is considerably smaller in the
  low-normalization, $\sigma_8=0.75$, simulation than in the
  $\sigma_8=1$ model. This is consistent with the picture that the
  anti-bias is caused by the halo disruption processes in clusters
  \citep{kravtsov_klypin99}, as the clusters in the low-$\sigma_8$
  model form later and the disruption processes have less time to
  operate.
  
\item The halo clustering strength depends on the maximum circular
  velocities of the halos (and hence their mass). The dependence is
  weak for $V_{\rm max}<200\ \rm km\, s^{-1}$, but becomes stronger
  for higher circular velocities.  The dependence of correlation
  length, $r_0$, on the number density of the halo sample is in
  general agreement with the clustering of galaxies in the SDSS survey.

\item We study the one- and two-halo components of the two-point
  correlation function of halos at different epochs. At $z=0$, the
  transition between these components is relatively smooth and the CF
  can be well fit by a single power law down to $\approx 100h^{-1}\ 
  \rm kpc$.  At higher redshifts the transition becomes more
  pronounced and occurs at smaller scales. The significant departures
  from the power law may thus be easier to detect in high-redshift
  galaxy surveys than at the present epoch. These departures can be used to
  put useful constraints on the environments and formation of
  galaxies.
  
\item If the deviations from a power law for the galaxy CF are as
  strong as indicated by our results, the assumption of the single
  power law often used in observational analyses of high-redshift
  clustering is dangerous and is likely to bias the estimates of the
  correlation length and/or slope of the correlation function.
  For the halos in our samples at $z=3$ the correlation function
  steepens at $r\sim 300h^{-1}$ comoving kpc.  There are indications
  that the clustering strength of $z=3$ LBGs becomes stronger than the
  large scale power-law at a similar scale \citep{adelberger_etal03}.
\end{itemize}

\acknowledgements 

We would like to thank Michael Blanton, Wayne Hu, Eduardo Rozo, 
Roman Scoccimarro, and David Weinberg for useful discussions and 
Ravi Sheth and Zheng Zheng for a careful
reading of the draft and many useful comments that improved
presentation.  This work was supported in part by the National Science
Foundation (NSF) under grants No.  AST-0206216 and AST-0239759, and,
in part, by NASA through grant NAG5-13274 and by the NSF Center
for Cosmological Physics at the University of Chicago. BA and JRP are
supported by NSF and NASA through grants NAG5-12326 and AST-0205944 to
the UCSC. AVK, AAB, and RHW, would like to thank Aspen Center for
Physics and organizers of the ``Cosmology and Astrophysics with Galaxy
Clusters'' workshop (June 2003), where this study was initiated, for
hospitality and productive environment. AAK and SG thank NSF/DAAD for
supporting their collaboration. The simulations and analyses presented
here were performed on the IBM RS/6000 SP3 system at the National
Energy Research Scientific Computing Center (NERSC) and on the
Origin2000 at the National Center for Supercomputing Applications
(NCSA).

\bibliography{hod}

\end{document}